\documentclass[letterpaper, conference]{IEEEtran}
\usepackage[utf8]{inputenc}
\usepackage[english]{babel}
\usepackage[ruled, vlined]{algorithm2e}
\usepackage{algpseudocode}
\usepackage{amssymb}
\usepackage{array}
\usepackage{arydshln}
\usepackage{blkarray, bigstrut}
\usepackage{bm}
\usepackage{booktabs}
\usepackage{braket}
\usepackage{cite}
\usepackage{cuted}
\usepackage{enumitem}
\usepackage{graphicx}
\usepackage[mathscr]{euscript}
\usepackage[switch]{lineno}
\usepackage{mathrsfs}
\usepackage{mathtools}
\usepackage{multirow}
\usepackage{multicol}
\usepackage{mathtools}
\usepackage{makecell}
\usepackage{xfrac}
\usepackage{framed} 
\usepackage[framed]{ntheorem}
\usepackage{pgfplots}
\usepackage{qtree}
\usepackage{adjustbox}
\usepackage{rotating}
\usepackage{float}
\usepackage{stfloats}
\usepackage{subcaption}
\usepackage{url}
\usepackage{tikz}
\usepackage{verbatimbox}
\usepackage[framemethod=TikZ]{mdframed}
\newframedtheorem{frm-thm}{}
\usetikzlibrary{patterns}
\usetikzlibrary{spy}
\usetikzlibrary{shapes, backgrounds,calc, snakes}
\usetikzlibrary{decorations.pathreplacing}
\usetikzlibrary{matrix}
\usepackage{fancybox}
\usepackage[normalem]{ulem}

\usepackage{balance}
\usepackage{lastpage}

\usepgfplotslibrary{fillbetween}
\usepgfplotslibrary{groupplots}
\pgfplotsset{compat=1.15,
        /pgfplots/ybar legend/.style={
        /pgfplots/legend image code/.code={%
        \draw[##1,/tikz/.cd,bar width=7pt, yshift=-0.45em, bar shift=1pt]
                plot coordinates {(0cm,0.8em)};
        },
},
}

\newcounter{groupcount}
\pgfplotsset{
    draw group line/.style n args={6}{
        after end axis/.append code={
            \setcounter{groupcount}{0}
            \pgfplotstableforeachcolumnelement{#1}\of\datatable\as\cell{%
                \def\temp{#2}
                \ifx\temp\cell
                    \ifnum\thegroupcount=0
                        \stepcounter{groupcount}
                        \pgfplotstablegetelem{\pgfplotstablerow}{X}\of\datatable
                        \coordinate [yshift=#4] (startgroup) at (axis cs:\pgfplotsretval,0);
                    \else
                        \pgfplotstablegetelem{\pgfplotstablerow}{X}\of\datatable
                        \coordinate [yshift=#4] (endgroup) at (axis cs:\pgfplotsretval,0);
                    \fi
                \else
                    \ifnum\thegroupcount=1
                        \setcounter{groupcount}{0}
                        \draw [
                            shorten >=-#6,
                            shorten <=-#6
                        ] (startgroup) -- node [anchor=base, yshift=0.5ex] {#3} (endgroup);
                    \fi
                \fi
            }
            \ifnum\thegroupcount=1
                        \setcounter{groupcount}{0}
                        \draw [
                            shorten >=-#6,
                            shorten <=-#6
                        ] (startgroup) -- node [anchor=base, yshift=0.5ex] {#3} (endgroup);
            \fi
        }
    }
}

\def\thesubsubsectiondis{\thesubsectiondis\arabic{subsubsection}}

\makeatletter
\def\subsubsection{%
  \@startsection
    {subsubsection}                 
    {3}                             
    {\parskip}                    
    {3.5ex plus 1.5ex minus 1.5ex}  
    {0.7ex plus .5ex minus 0ex}     
    {\normalfont\normalsize\itshape}
}
\makeatother

\newcommand{\argmax}{\arg\!\max}

\definecolor{bblue}{rgb}{0.12392, 0.0490, 0.9588}
\definecolor{sskyblue}{rgb}{0.1529, 0.5882, 0.9216}
\definecolor{ggreen}{rgb}{0.7098, 0.95, 0.40781}
\definecolor{yyellow}{rgb}{0.9765, 0.9804, 0.0784}

\definecolor{color6}{HTML}{03B3FF}
\pgfplotsset{major grid style={densely dotted,gray!50!gray}}
\definecolor{dcolor1}{HTML}{253494}
\definecolor{dcolor2}{HTML}{636363}

\definecolor{dcolor3}{HTML}{fcbba1}
\definecolor{dcolor4}{HTML}{fb6a4a}
\definecolor{dcolor5}{HTML}{cb181d}
\definecolor{dcolor6}{HTML}{67000d}

\definecolor{dcolor7}{HTML}{9ecae1}
\definecolor{dcolor8}{HTML}{4292c6}
\definecolor{dcolor9}{HTML}{08519c}

\definecolor{gcolor7}{HTML}{272727}

\definecolor{lcolor1}{HTML}{272727}
\definecolor{lcolor2}{HTML}{053061}
\definecolor{lcolor3}{HTML}{4393c3}

\definecolor{lcolor4}{HTML}{d6604d}
\definecolor{lcolor5}{HTML}{b2182b}
\definecolor{lcolor6}{HTML}{67001f}
\definecolor{lcolor7}{HTML}{272727}

\makeatletter
\def\newmaketag{%
	\def\maketag@@@##1{\hbox{\m@th\normalfont\normalsize##1}}%
}
\makeatother

\usepackage[textsize=scriptsize]{todonotes}

\DeclareMathSizes{10pt}{8pt}{6pt}{5pt} 	

\begin{document}
\title{\LARGE{BEAMWAVE: Cross-layer Beamforming and Scheduling for Superimposed Transmissions in Industrial IoT mmWave Networks}}

\author{
\IEEEauthorblockN{Luis F. Abanto-Leon, and Matthias Hollick, and Gek Hong (Allyson) Sim} \\ 
\IEEEauthorblockA{Secure Mobile Networking Lab, Technische Universit\"{a}t Darmstadt, Germany} \\ \{labanto, mhollick, asim\}@seemoo.tu-darmstadt.de
}

\markboth{Journal of \LaTeX\ Class Files,~Vol.~33, No.~27, Jun~2019}%
{Shell \MakeLowercase{\textit{et al.}}: Bare Demo of IEEEtran.cls for IEEE Communications Society Journals}

\maketitle

\begin{abstract}
The omnipresence of IoT devices in Industry 4.0 is expected to foster higher reliability, safety, and efficiency. However, interconnecting a large number of wireless devices without jeopardizing the system performance proves challenging. To address the requirements of future industries, we investigate the cross-layer design of beamforming and scheduling for layered-division multiplexing (LDM) systems in millimeter-wave bands. Scheduling is crucial as the devices in industrial settings are expected to proliferate rapidly. Also, highly performant beamforming is necessary to ensure scalability. By adopting LDM, multiple transmissions can be non-orthogonally superimposed. Specifically, we consider a \emph{superior-importance control multicast message} required to be ubiquitous to all devices and \emph{inferior-importance private unicast messages} targeting a subset of scheduled devices. Due to NP-hardness, we propose \texttt{BEAMWAVE}, which decomposes the problem into \emph{beamforming} and \emph{scheduling}. Through simulations, we show that \texttt{BEAMWAVE} attains near-optimality and outperforms other competing schemes.
\end{abstract}


\IEEEpeerreviewmaketitle

\vspace{-4mm}
\section{Introduction} \label{introduction}
Industry 4.0 envisions automated factories with a massive number of interconnected industrial internet-of-things (IoT) devices \cite{chen2019:massive-access-iot}, such as sensors, actuators, programmable logic devices, and access points. Such degree of interconnectivity is expected to facilitate ultra-precise control and seamless coordination, thus enabling extremely efficient and dependable manufacturing processes \cite{abanto2020:hydrawave-multigroup-multicast-hybrid-precoding-scheduling-industry}. In the existing industrial settings, the majority of stationary devices are interconnected through redundant wired connections to guarantee communications with high reliability. However, with the upsurge of devices in smart industries, wired solutions will encounter the following problems: \emph{(i)} intricate implementation complexity to interconnect a massive number of devices, \emph{(ii)} increased operational costs due to hard-wiring, \emph{(iii)} limited maneuverability of articulated robots, and \emph{(iv)} communication infeasibility with autonomous mobile freight transport. In contrast, wireless solutions can substantially simplify the deployment complexity and reduce maintenance costs while promoting the adoption of more flexible mechanics and mobile apparatus. \emph{Thus, the transformation from wired to wireless infrastructure is an appealing strategy towards the evolution of industries.}

By harnessing \emph{millimeter-wave} (mmWave) and \emph{massive multiple-input multiple-output} (mMIMO), high spectral efficiency has been demonstrated (e.g., \cite{bjornson2019:sub6ghz-vs-mmwave, ribeiro2018:energy-efficiency-mmwave-mimo}). Specifically, mmWave is an attractive substitute for the saturated sub-6 GHz spectrum due to broad bandwidth availability. Also, because of the shorter wavelength, mmWave requires miniature antennas that can be easily embedded onto small industrial devices. Further, mmWave exhibits high spatial reuse due to severe path-loss and sparse propagation, making it ideal for short-range communications in extremely dense scenarios such as the industrial settings. Besides, owing to increased degrees of freedom, mMIMO renders extraordinary interference mitigation \cite{larsson2014:massive-mimo-next-generation, bjornson2018:massive-mimo-capacity} that enables augmented spectral efficiency and exceptional multiplexing capability, which are desirable features to support the future industrial landscape. 


In factories of the future, industrial devices will require two types of information: \emph{shared safety/control messages} (multicast signal) and \emph{private messages} (unicast signals). Such a requirement could be addressed by orthogonal multiple access (OMA) schemes, wherein multicast and unicast signals would be transmitted in disjoint time or frequency resources. Nevertheless, with the anticipated escalation, OMA schemes will struggle to accommodate a large number of devices in orthogonal resources. Thus, \emph{non-orthogonal multiple access} (NOMA) schemes are envisaged as a remedy to cope with the scarcity of radio resources. In particular, NOMA can boost the spectral efficiency by admitting superposed transmissions in the power or code domain. Among the plethora of NOMA variants \cite{liu2017:noma-5g-beyond}, layered-division multiplexing (LDM) has been recognized as a promising candidate to meet the growing spectrum demands. LDM is a power-domain NOMA scheme capable of conveying multiple layers of information simultaneously while using the same time-frequency resources. By harnessing LDM in industrial settings, multicast and unicast information can be disseminated concurrently without resorting to OMA schemes such as time/frequency-division multiplexing (T/FDM).

Several NOMA schemes have recently been intertwined with mmWave and mMIMO, showing remarkable synergy in many use cases (e.g., \cite{zhao2016:ldm-broadcast-unicast, chen2018:nonorthogonal-massive-access, ding2017:random-beamforming-noma}). Also, preliminary studies on the usage of NOMA \cite{3gpp2020:TR21.916} and mmWave \cite{loch2019:measurement-campaign-mmwave} for smart industries have shown favorable results. Based on this evidence, it is expected that by jointly leveraging mmWave, mMIMO and LDM, the stringent requirements of future industrial ecosystems can be fulfilled. However, the synthesis of these technologies poses challenges that require further study when considered in the context of Industry 4.0.

\noindent\textbf{Challenges:} The following summarizes relevant aspects that need to be considered in the envisaged industrial landscape.

\begin{itemize} [leftmargin=0.3cm]
	\item The maximum number of devices that can be simultaneously served with individual signals is limited by the number of radio frequency (RF) chains at the transmitter (e.g., base station). Hence, with the forecasted rapid escalation of devices in industrial sectors \cite{palattella2016:iot-5g-architectures}, the problem aggravates. Most existing works on beamforming consider sufficient RF chains to serve all devices, thus rendering scheduling unnecessary. \textit{However, as networks densify, scheduling will be pivotal in exerting substantial improvement in the system performance. Thus, considering the cross-layer optimization of beamforming and scheduling is of utmost importance.}

	\item Multicast and unicast transmissions give rise to conflicting objectives. From the multicast perspective, the transmitter consumes lesser power while the spectral efficiency improves when the devices have correlated channels. From the unicast perspective, we observe the opposite effect, i.e., correlated channels yield low spectral efficiency while demanding higher power. \textit{As a result, selecting a suitable set of devices (i.e., scheduling) in superimposed multicast-unicast LDM systems requires special consideration.}
	
	\item Problems dealing with cross-layer optimization of beamforming and scheduling are challenging to solve due its inherent nature of involving integer and continuous variables.
\end{itemize}

\noindent\textbf{Research problem:} Due to safety reasons, the superior-importance multicast signal (e.g., control messages) is not subject to scheduling but is required to be ubiquitous to all IoT devices. Contrastingly, the inferior-importance unicast signals (e.g., software updates) are conveyed to only a specific subset of devices (i.e., scheduling) subject to RF chains availability. As a result, two superimposed beamformers are designed. One beamformer transmits the control signal to all devices. The second beamformer caters a selected subset of devices with private unicast signals, where the selection of devices is inspired by the max-min criterion.

\noindent\textbf{Related work:} Beamforming in LDM systems has been studied for \emph{(i)} transmit power minimization \cite{zhao2016:ldm-broadcast-unicast, zhao2019:ldm-broadcast-unicast, liu2017:joint-multicast-unicast-beamforming-miso}, \emph{(ii)} energy efficiency improvement \cite{li2019:energy-efficient-noma-multicast-unicast}, \emph{(iii)} joint beamforming and base station clustering \cite{chen2017:backhaul-constrained-beamforming-multicast-unicast, chen2018:bs-clustering-beamforming-ldm-backhaul}, \emph{(iv)} sum-rate maximization \cite{wang2018:hybrid-beamforming-joint-unicast-multicast}, \emph{(v)} simultaneous wireless information and power transfer (SWIPT) \cite{hao2019:beamforming-swipt-ldm-lens-array, hao2019:energy-efficient-hybrid-precoding-swipt}, and \emph{(vi)} fairness improvement \cite{abanto2020:fairness-hybrid-precoding-unicast-multicast}. \emph{To the best of our knowledge, the cross-layer optimization problem for joint design of beamforming and scheduling in LDM systems has not been studied before. Further, the combination of mmWave, mMIMO and LDM has neither been studied in industrial settings.}

\noindent\textbf{Contributions: } Our contributions are the following. 

\begin{itemize} [leftmargin=0.3cm]
	\item We formulate a NP-hard problem ($ \mathcal{P} $) that jointly optimizes \emph{beamforming} and \emph{scheduling} for multicast-unicast LDM transmissions, where we impose a signal-to-interference-plus-noise ratio (SINR) constraint on the multicast signal to ensure that every IoT device correctly decodes the ubiquitous safety message. 
	
	\item To solve problem $ \mathcal{P} $ we propose \texttt{BEAMWAVE}, which decomposes $ \mathcal{P} $ into two problems $ \mathcal{S} $ and $ \mathcal{D} $.  We propose a novel scheduling scheme $ \mathcal{S} $ based on new pair-wise metrics, \texttt{PAWN}, \texttt{ROOK}, \texttt{KING}, that we devise to guide the decision. Essentially, these metrics represent the discordance of co-scheduling two devices together. To solve $ \mathcal{D} $, we devise an approach based on the convex-concave procedure (CCP). Through simulations, we show that the proposed \texttt{BEAMWAVE} can attain near-optimality when compared to an exhaustive search approach.
	
	\item We motivate the need for scheduling in LDM systems, specially when the number of RF chains is insufficient to serve a significantly larger number of devices (which is expected in future industrial settings). In addition, we apply our proposed scheduler $ \mathcal{S} $ to T/FDM systems to find the set of devices co-scheduled in the same time or frequency resource. Through simulations, we show the importance of scheduling when compared to more trivial schemes such as random selection.
\end{itemize}


\section{System Model} \label{system_model}

We assume a system, where a next-generation Node B (gNodeB) serves $ K $ devices indexed by $ \mathcal{K} = \left\lbrace 1, \cdots, K \right\rbrace $. The gNodeB transmits a signal composed of two non-orthogonal layers. The \textit{primary layer} is a multicast signal that conveys a shared control message intended for every device $ k \in \mathcal{K} $. The \emph{secondary layer} is a composite signal consisting of multiple unicast messages intended for a subset of devices $ \mathcal{K}' \subseteq \mathcal{K} $, where $ K' = \left| \mathcal{K}' \right| $. Thus, $ K' $ \emph{dual-layer} devices are catered with simultaneous unicast and multicast transmissions, whereas $ K - K' $ \emph{single-layer} devices are served with multicast information only. The gNodeB possesses a precoder (i.e., \textit{transmit beamformer}) consisting of $ N_\mathrm{tx} $ antennas and $ N^\mathrm{RF}_\mathrm{tx} << N_\mathrm{tx} $ RF chains. Without loss of generality, we assume that $ N^\mathrm{RF}_\mathrm{tx} = K' $. Besides, each IoT device in the system is equipped with a single RF chain (i.e., $ N^\mathrm{RF}_\mathrm{rx} = 1 $) and $ N_\mathrm{rx} $ antennas. 

The downlink signal from the gNodeB is denoted by $\mathbf{x} = \left[ \mathbf{B} \vert \mathbf{m} \right] \left[ \mathbf{s}^T \vert z \right]^T $. The unicast and multicast precoders are represented by $ \mathbf{B} \in \mathbb{C}^{N_\mathrm{tx} \times K'} $ and $ \mathbf{m} \in \mathbb{C}^{N_\mathrm{tx} \times 1}$, respectively. In addition, $ \mathbf{s} \in \mathbb{C}^{K' \times 1} $ denotes the unicast symbols for the \emph{dual-layer} devices while $ z \in \mathbb{C} $ is the shared multicast symbol intended for all $ K $ devices, with $ \mathbb{E} \left\lbrace \left[ \mathbf{s}^T, z \right]^H \left[ \mathbf{s}^T, z \right] \right\rbrace = \mathbf{I} $. More specifically, $ \mathbf{B} = \widetilde{\mathbf{B}} \mathbf{U} $ where $ \widetilde{\mathbf{B}} = \left[ \mathbf{b}_1, \dots, \mathbf{b}_{K} \right] \in \mathbb{C}^{N_\mathrm{tx} \times K} $ and $ \mathbf{U} \in \mathbb{B}^{K \times K'} $ is a binary matrix. Also, $ \mathbf{s} = \mathbf{U}^T \widetilde{\mathbf{s}} $ where $ \widetilde{\mathbf{s}} = \left[ s_1, \dots, s_K \right]^T \in \mathbb{C}^{K \times 1}  $. Concretely, the matrix $ \mathbf{U} $ selects the \emph{dual-layer} devices that will be served with both unicast and multicast signals. Thus, it must hold that $ \mathbf{1}^T \mathbf{U} \mathbf{1} = K' $, $ \mathbf{U} \mathbf{1} \preccurlyeq \mathbf{1} $ and $ \mathbf{U}^T \mathbf{1} \preccurlyeq \mathbf{1} $. As a result, $ \mathbf{U} \mathbf{U}^T = \mathrm{diag} \left( \left[ \mu_1, \cdots, \mu_{K} \right] \right) $ is a square matrix whose $ k $-th diagonal element is $ 1 $ when $ k $ is a \emph{dual-layer} device (i.e., $ \mu_k = \left[ \mathbf{U} \mathbf{U}^T \right]_{k,k} = 1 $, if $ k \in \mathcal{K}' $). Otherwise, $ \mu_k = \left[ \mathbf{U} \mathbf{U}^T \right]_{k,k} = 0 $ when $ k $ is a \emph{single-layer} device. Assuming flat fading, the signal received by device $ k \in \mathcal{K} $ is given by
\begin{align} \label{e1}
	\begin{split}
		y_k & = \underbrace{\mathbf{w}^H_k \mathbf{H}_k \mathbf{m} z}
						_{ y^\mathrm{M}_k : \text{ multicast signal}}
						+ \underbrace{\mathbf{w}^H_k \mathbf{H}_k \sum_{j \in \mathcal{K}'}{\mathbf{b}_j s_j}}
						_{ y^\mathrm{U}_k : \text{ aggregate unicast signal}}		
						+ \underbrace{\mathbf{w}^H_k \mathbf{n}_k,}_{\eta_k: \text{noise}} \\			
	\end{split}
\end{align}
where $ \mathbf{w}^H_k \mathbf{H}_k \sum_{j \in \mathcal{K} \backslash \mathcal{K}'}{ \mu_j \mathbf{b}_j s_j} = 0 $ since $ \mu_j = 0 $, $ \forall j \in \mathcal{K} \backslash \mathcal{K}' $. Besides, $\mathbf{w}_k \in {\mathbb{C}}^{N_\mathrm{rx} \times 1}$ represents the combiner (i.e., \textit{receive beamformer}) of the $ k $-th device, $\mathbf{n}_k \sim \mathcal{CN} \left( \mathbf{0}, {\sigma}^2 \mathbf{I} \right) $ symbolizes circularly symmetric Gaussian noise whereas $\mathbf{H}_k \in {\mathbb{C}}^{N_\mathrm{rx} \times N_\mathrm{tx}}$ denotes the channel between the gNodeB and the $ k $-th device, defined as

\begin{align} \label{e1a}
	\mathbf{H}_k = \sqrt{\frac{N_\mathrm{rx} N_\mathrm{tx} }{L_k}}\sum_{l=1}^{L_k} \rho^{(l)}_k \mathbf{a}_\mathrm{rx} \left( \psi^{(l)}_k \right)  \mathbf{a}_\mathrm{tx} \left( \phi^{(l)}_k \right)^H. 
\end{align}

Here, $ L_k $ is the number of paths in $ \mathbf{H}_k $, whereas $ \psi^{(l)}_k $ and $ \phi^{(l)}_k $ represent the angle of arrival (AoA) and angle of departure (AoD) of the $ l $-th path in $ \mathbf{H}_k $, respectively. The array vector responses at the $ k $-th device and gNodeB, in the directions of $ \psi^{(l)}_k $ and $ \phi^{(l)}_k $, are respectively defined as $ \mathbf{a}_\mathrm{rx} \left( \psi^{(l)}_k \right) = \frac{1}{\sqrt{N_\mathrm{rx}}} \left[ 1, \cdots, e^{-j (N_\mathrm{rx} - 1) \frac{2 \pi}{\lambda} d \cos (\psi^{(l)}_k)} \right]^T $ and $ \mathbf{a}_\mathrm{tx} \left( \phi^{(l)}_k \right) = \frac{1}{\sqrt{N_\mathrm{tx}}} \left[ 1, \cdots, e^{-j (N_\mathrm{tx} - 1) \frac{2 \pi}{\lambda} d \cos (\phi^{(l)}_k)} \right]^T $. Also, $ \frac{d}{\lambda} = 0.5 $ and $ \rho^{(l)}_k $ is the complex gain of the $ l $-th path in $ \mathbf{H}_k $, which is represented as a random variable following a complex Gaussian distribution $ \mathcal{C} \mathcal{N} (0, 1) $.

Due to the superposed structure of the transmitted signal, successive interference cancellation (SIC) is performed by the \emph{dual-layer} devices in order to extract multicast and unicast information. Every device $ k \in \mathcal{K} $ decodes the multicast symbol first by treating the aggregate unicast signal as noise. In addition, if $ k $ is a \emph{dual-layer} device (i.e., $ k \in \mathcal{K}' $), then the device applies SIC decoding. Essentially, the $ k $-th device reconstructs the multicast signal $ y^\mathrm{M}_k $ using the decoded symbol $ z $, and then subtracts $ y^\mathrm{M}_k $ from $ y_k $. Thereupon, the remaining byproduct consists solely of unicast components ($ y^\mathrm{U}_k $) and noise ($ \eta_k $), from where the \emph{dual-layer} device can decode its intended symbol $ s_k $. The SINR of the multicast and unicast signals at the $ k $-th device are respectively defined as
\begin{equation} \label{e2}
	\mathrm{SINR}^\mathrm{M}_k = \frac{\left| \mathbf{w}^H_k \mathbf{H}_k \mathbf{m} \right|^2 }
					   		{
			                  \sum_{j \in \mathcal{K}'} \left| \mathbf{w}^H_k \mathbf{H}_k {\mathbf{b}_j} \right|^2 + {\sigma}^2 \left\| \mathbf{w}_k \right\|^2_2 
		           	   		}, \forall k \in \mathcal{K},
\end{equation}
\begin{equation} \label{e3}
	\mathrm{SINR}^\mathrm{U}_k = \frac{\left| \mathbf{w}^H_k \mathbf{H}_k {\mathbf{b}_k} \right|^2 }
			   			{
							\sum_{j \neq k, j \in \mathcal{K}'} \left| \mathbf{w}^H_k \mathbf{H}_k {\mathbf{b}_j} \right|^2 + {\sigma}^2 \left\| \mathbf{w}_k \right\|^2_2
			   			}, \forall k \in \mathcal{K}'.
\end{equation}

\section{Problem Formulation} \label{problem_formulation}
We present a joint formulation that encompasses the optimization of \emph{(i)} scheduling, \emph{(ii)} precoders and \emph{(iii)} combiners,
\begin{subequations} \label{e4}
	\begin{align}
		\mathcal{P}: & \max_{
				\substack{
							\mathbf{W}, \mathbf{m}, \\
							\widetilde{\mathbf{B}}, \boldsymbol{\mu}
						 }
			   } 
		& & \min_{k \in \mathcal{K}} 
								\frac{\left| \mathbf{w}^H_k \mathbf{H}_k {\mathbf{b}_k} \right|^2 g(\mu_k)}
				 				{ \sum_{ j \neq k, j \in \mathcal{K} } \left| \mathbf{w}^H_k \mathbf{H}_k {\mathbf{b}_j} \right|^2 \mu_j + \sigma^2 \left\| \mathbf{w}_k \right\|^2_2 } \nonumber		
		\\
		\vspace{-0.2cm}
		& ~~ \mathrm{s.t.} & \mathrm{C_1}: ~ & \frac{\left| \mathbf{w}^H_k \mathbf{H}_k \mathbf{m} \right|^2 }
		{ \sum_{j \in \mathcal{K}} \left| \mathbf{w}^H_k \mathbf{H}_k {\mathbf{b}_j} \right|^2 \mu_j + {\sigma}^2 \left\| \mathbf{w}_k \right\|^2_2 } \geq \gamma_\mathrm{min}, \forall k \in \mathcal{K}, \nonumber
		\\
		& & \mathrm{C_2}: ~ & \sum_{k \in \mathcal{K}} \left\| \mathbf{b}_k \right\|^2_2 \mu_k + \left\| \mathbf{m} \right\|^2_2 \leq P_\mathrm{tx}, \nonumber
		\\ 
		& & \mathrm{C_3}: ~ & \sum_{k \in \mathcal{K}} \mu_k = K', \nonumber
		\\ 
		& & \mathrm{C_4}: ~ & \left[ \mathbf{w}_k \right]_l \in \mathcal{W}, l \in \mathcal{L}, \forall k \in \mathcal{K}, \nonumber
		\\ 
		& & \mathrm{C_5}: ~ & \mu_k \in \left\lbrace 0, 1\right\rbrace, \nonumber
	\end{align}
\end{subequations}
where $ g(\chi) $ is defined as
\begin{align} \nonumber
		g(\chi) = 
				 	\begin{cases}
				      1, 	  & \text{if} ~ \chi = 1, \\
				      \infty, & \text{if} ~ \chi = 0.
				    \end{cases} 
\end{align}
and $ \mathbf{W} = \left[ \mathbf{w}_1, \cdots, \mathbf{w}_K \right] $, $ \widetilde{\mathbf{B}} = \left[ \mathbf{b}_1, \cdots, \mathbf{b}_K \right] $, $ \boldsymbol{\mu} = \left[ \mu_1, \cdots, \mu_K \right] $.

The objective function of $ \mathcal{P} $ aims to find the subset $ \mathcal{K}' \subseteq \mathcal{K} $ that maximizes the minimum $ \mathrm{SINR}^\mathrm{U}_k $, $ k \in \mathcal{K}' $. The constraint $ \mathrm{C_1} $ requires $ \mathrm{SINR}^\mathrm{M}_k $ to be above a threshold $ \gamma_\mathrm{min} $ for all devices, whereas $ \mathrm{C_2} $ limits the transmit power to $ P_\mathrm{tx} $. The constraint $ \mathrm{C_3} $ selects $ K' $ devices for \emph{dual-layer} transmissions while $ \mathrm{C_4} $ enforces beamforming restrictions on the combiners. Specifically, only a small number of $ L_\mathrm{rx} $ constant-modulus phase shifts are admitted for designing the combiners. Every phase shift $ \left[ \mathbf{w}_k \right]_l $ is confined to $ \mathcal{W} = \left\lbrace \delta_\mathrm{rx}, \dots, \delta_\mathrm{rx} e^{j \frac{2 \pi \left( L_\mathrm{rx} - 1 \right)}{L_\mathrm{rx}}} \right\rbrace $, $ l \in \mathcal{L} = \left\lbrace 1, \dots, N_\mathrm{rx} \right\rbrace $\footnote{Realize that $ \mathcal{W} $ consists of equally-distributed phase rotations with magnitude $ \delta_\mathrm{rx} $, where $ L_\mathrm{rx} $ defines the phase resolution.}. Finally, $ \mathrm{C_5} $ enforces the Boolean nature of $ \mu_k $. We consider limited receive power $ P_\mathrm{rx} $ at each device. Thus, $ P_\mathrm{rx} = \left\| \mathbf{w}_k \right\|^2_2 = \sum^{N_\mathrm{rx}}_{l = 1} \left| \left[ \mathbf{w}_k \right]_l \right|^2 = N_\mathrm{rx} \delta^2_\mathrm{rx} $, where $ \delta_\mathrm{rx} = \sqrt{P_\mathrm{rx} / N_\mathrm{rx}} $.

\begin{figure}[!t]
		\centering
	    \includegraphics[width = 0.85\columnwidth]{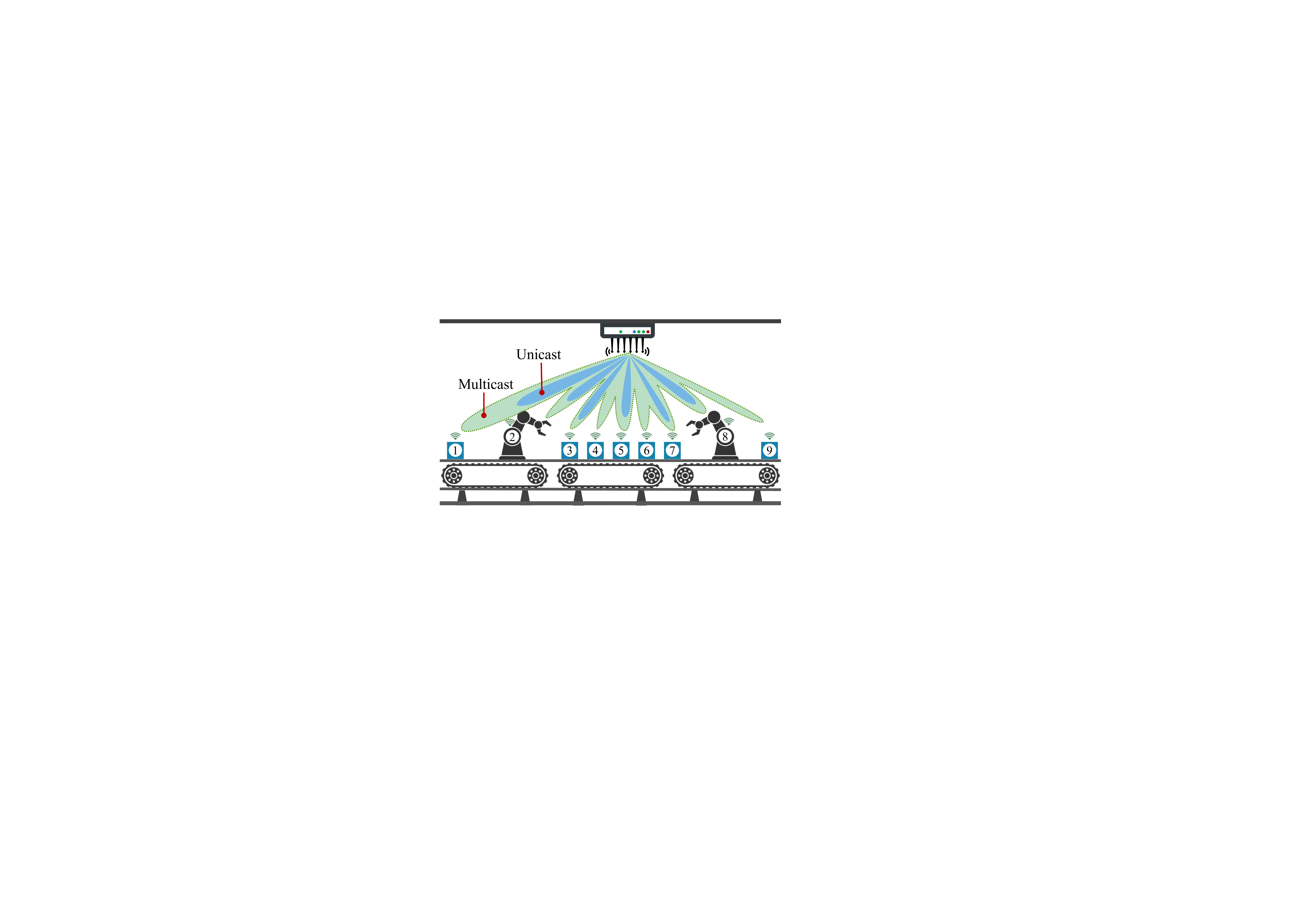}
		\caption{Multicast-unicast LDM system with $ K = 9 $ devices. \emph{The multicast signal is intended for all devices whereas only a subset of $ K' = 6 $ devices is served with private unicast signals.}}
		\label{f1}	
		\vspace{-2mm}
\end{figure}

To solve $ \mathcal{P} $, one possibility is to adopt an exhaustive search approach ($ \texttt{XHAUS} $). This procedure consists in generating every subset of devices of size $ K' $ from a total of $ K $, thus yielding $ J = { K \choose K'} $ possibilities for $ \boldsymbol{\mu} $, i.e., $ \left\lbrace \boldsymbol{\mu}_1, \cdots, \boldsymbol{\mu}_J \right\rbrace $. Then, $ \mathcal{P} $ is solved for each of the combinations, i.e., $ \left\lbrace  \mathcal{P} \left( \mathbf{W}, \mathbf{m}, \widetilde{\mathbf{B}}, \boldsymbol{\mu}_1 \right), \cdots, \mathcal{P} \left( \mathbf{W}, \mathbf{m}, \widetilde{\mathbf{B}}, \boldsymbol{\mu}_J \right) \right\rbrace $ and the choice that attains the \emph{max-min} unicast SINR is selected as optimal. While $ \texttt{XHAUS} $ yields the best scheduling, it is computationally expensive. Therefore, in Section \ref{relaxed_formulation}, we propose a scheme, wherein $ \boldsymbol{\mu} $ is determined in advance by a novel scheduler. Then, $ \mathbf{W}, \mathbf{m}, \widetilde{\mathbf{B}} $ are designed for the resulting selection of devices\footnote{Notice that even for a given $ \boldsymbol{\mu}' $, the problem $ \mathcal{P} \left( \mathbf{W}, \mathbf{m}, \widetilde{\mathbf{B}}, \boldsymbol{\mu}' \right) $ is nonconvex and challenging to solve.}. Problem $ \mathcal{P} $ is illustrated in Fig. \ref{f1}. 

\section{\texttt{BEAMWAVE}: Proposed Scheme} \label{relaxed_formulation}
We divide $ \mathcal{P} $ into two problems: $ \mathcal{S} $ (Section \ref{scheduling}) and $ \mathcal{D} $ (Section \ref{precoder_combiners}). First, $ \mathcal{S} $ finds a subset $ \mathcal{K}' $ of \emph{dual-layer} devices, thus rendering the binary scheduling variables available. Subsequently, $ \mathcal{D} $ designs the precoder and the combiners.

\subsection{Scheduling} \label{scheduling}
Selecting an optimal subset of \emph{dual-layer} devices $ \mathcal{K}' $ that leads to the maximization of the minimum unicast SINR is intrinsically of combinatorial nature. In order to circumvent the exhaustive search, we propose a novel scheduling scheme $ \mathcal{S} $, which is based on the minimization of an aggregate pairwise device-specific channel metric. The objective is to find the variables $ \boldsymbol{\mu} $ and $ \boldsymbol{\nu} $ such that $ f_\mathcal{S} \left( \boldsymbol{\nu} \right) $ is minimized. 
\begin{subequations}
	\begin{align}
	\mathcal{S}: & \min_{ \boldsymbol{\mu}, \boldsymbol{\nu} } & & f_\mathcal{S} \left( \boldsymbol{\nu} \right) \triangleq \sum^{K - 1}_{j = 1} \sum^{K}_{l = j + 1} \theta_{j,l} \cdot \nu_{j,l} \nonumber
	\\
	\vspace{-0.2cm}
	& ~ \mathrm{s.t.} & \mathrm{Q_1}: ~ & \mu_j \geq \nu_{j,l}, \forall j < l, \nonumber
	\\
	& & \mathrm{Q_2}: ~ & \mu_j + \mu_l \leq 1 +\nu_{j,l}, \forall j < l, \nonumber
	\\
	& & \mathrm{Q_3}: ~ & \sum^K_{j = 1} \mu_j = K', \nonumber
	\\
	& & \mathrm{Q_4}: ~ & \mu_j \in \left\lbrace 0, 1 \right\rbrace, \forall j, \nonumber
	\\
	& & \mathrm{Q_5}: ~ & \nu_{j,l} \in \left\lbrace 0, 1 \right\rbrace, \forall j < l. \nonumber
	\end{align}
\end{subequations}

In particular, $ \theta_{j,l} $ denotes a positive metric between two devices $ j \in \mathcal{K} $ and $ l \in \mathcal{K} $, representing the discordance of co-scheduling the two devices. The auxiliary variable $ \nu_{j,l} $, assumes the value of $ 1 $, if devices $ j $ and $ l $ are co-scheduled for \emph{dual-layer} transmissions. Otherwise, $ \nu_{j,l} = 0 $. As defined in $ \mathcal{P} $, the variable $ \mu_j $ denotes with $ 1 $ that $ j \in \mathcal{K} $ is a \emph{dual-layer} device. The constraints $ \mathrm{Q_1} $ and $ \mathrm{Q_2} $ have been included in order to bind the two sets of variables, i.e., $ \boldsymbol{\mu} $ and $ \boldsymbol{\nu} $. Specifically, $ \mathrm{Q_1} $ states that $ \nu_{j,l} $ is upper-bounded by $ \mu_j $ since $ \nu_{j,l} $ can only be $ 1 $ when the devices $ j $ and $ l $ are co-scheduled. Similarly, $ \mathrm{Q_2} $ is a lower bound for $ \nu_{j,l} $ in terms of $ \mu_j $ and $ \mu_l $. Besides, $ \mathrm{Q_3} $ restricts the maximum number of \emph{dual-layer} devices to $ K' $. Constraints $ \mathrm{Q_4} - \mathrm{Q_5} $ denote the Boolean nature of the variables.

We denote the solution of $\mathcal{S} $ by $ \left( \boldsymbol{\mu}^\star, \boldsymbol{\nu}^\star \right)  $. In the following, we propose three metrics $ \theta_{j,l} $ (i.e., \texttt{PAWN}, \texttt{ROOK}, \texttt{KING}), based on channel correlation and channel energy, which will support the scheduling decision.

\vspace{1mm}
{\texttt{{CORR:}}} Channel correlation has been extensively used for multiuser unicast scheduling in prior literature (e.g., \cite{yoo2007:multi-antenna-downlink-user-selection}). Given any two devices $ j $ and $ l $, \texttt{CORR} is computed as $ \theta_{j,l} = \frac{ \left| \mathbf{h}^H_j \mathbf{h}_l \right| }{ \left\|  \mathbf{h}_j \right\|_2  \left\|  \mathbf{h}_l \right\|_2  } $, where $ \mathbf{h}_j = \mathrm{vec} \left( \mathbf{H}_j \right) $. Intuitively, a large value of $ 0 \leq \theta_{j,l} \leq 1 $ implies that the two devices have correlated channels and therefore they are prone to generate more interference to each other. \texttt{CORR} has conventionally been used in a greedy manner, where users/devices are sequentially chosen based on the cumulative correlation with respect to the already selected devices. In contrast, herein we use \texttt{CORR} in combination with our proposed scheduler $ \mathcal{S} $, thus allowing to find the best set $ \mathcal{K}' $ of \textit{dual-layer} devices that renders the least aggregate pair-wise channel correlation in the sense of $ f_\mathcal{S} \left( \boldsymbol{\nu} \right) $.

\vspace{1mm}
{\texttt{{PAWN:}}} We propose this metric as a generalization of \texttt{CORR}, where we compute the channel correlation between all the rows of $ \mathbf{H}_j $ and $ \mathbf{H}_l $. For two devices $ j $ and $ l $, the metric is expressed as $ \theta_{j,l} = \sum^{N_\mathrm{rx}}_{n_1 = 1} \sum^{N_\mathrm{rx}}_{n_2 = 1} \frac{1}{N^2_\mathrm{rx}} \frac{ \left| \mathbf{H}_j (n_1) \mathbf{H}^H_l (n_2) \right| }{ \left\|  \mathbf{H}_j (n_1) \right\|_2  \left\|  \mathbf{H}_l (n_2) \right\|_2 } $, with $ \mathbf{H}_j (n) $ denoting the $ n $-th row of $ \mathbf{H}_j $. Note that for the special case of $ N_\mathrm{rx} = 1 $, \texttt{CORR} and \texttt{PAWN} are equivalent.

\vspace{1mm}
{{\texttt{ROOK}:}} We devise this metric as a combination of two components. One of the constituents leverages the channel energy difference between two devices. The second component is the metric \texttt{PAWN}. Thus, \texttt{ROOK} is defined as $ \theta_{j,l} = \omega \frac{\left| \left\| \mathbf{H}_j \right\|^2_\mathrm{F} - \left\| \mathbf{H}_l \right\|^2_\mathrm{F} \right| }{ \left\| \mathbf{H}_j \right\|^2_\mathrm{F} + \left\| \mathbf{H}_l \right\|^2_\mathrm{F} } + (1 - \omega) \sum^{N_\mathrm{rx}}_{n_1 = 1} \sum^{N_\mathrm{rx}}_{n_2 = 1} \frac{1}{N^2_\mathrm{rx}} $ $ \frac{ \left| \mathbf{H}_j (n_1) \mathbf{H}^H_l (n_2) \right| }{ \left\|  \mathbf{H}_j (n_1) \right\|_2 \left\| \mathbf{H}_l (n_2) \right\|_2 } $ with $ 0 \leq \omega \leq 1 $. The rationale for this metric is that devices with uncorrelated channel vectors and comparable channel energy are desirable for scheduling.

\vspace{1mm}
{{\texttt{KING}:}} We also devise this metric as a combination of two components. Specifically, we combine \texttt{PAWN} with the ratio between the channel energy of a device and the largest channel energy among all the devices. Thus, $ \theta_{j,l} = \omega \left( \frac{\left\| \mathbf{H}_\mathrm{max} \right\|^2_\mathrm{F} - \left\| \mathbf{H}_j \right\|^2_\mathrm{F} }{ \left\| \mathbf{H}_\mathrm{max} \right\|^2_\mathrm{F} } + \frac{\left\| \mathbf{H}_\mathrm{max} \right\|^2_\mathrm{F} - \left\| \mathbf{H}_l \right\|^2_\mathrm{F} }{ \left\| \mathbf{H}_\mathrm{max} \right\|^2_\mathrm{F} } \right) + (1 - \omega) \sum^{N_\mathrm{rx}}_{n_1 = 1} \sum^{N_\mathrm{rx}}_{n_2 = 1} \frac{1}{N^2_\mathrm{rx}} $ $ \frac{ \left| \mathbf{H}_j (n_1) \mathbf{H}^H_l (n_2) \right| }{ \left\|  \mathbf{H}_j (n_1) \right\|_2 \left\| \mathbf{H}_l (n_2) \right\|_2 } $, where $ \left\| \mathbf{H}_\mathrm{max} \right\|^2_\mathrm{F} = \max_{j \in \mathcal{K}} \left\| \mathbf{H}_j \right\|^2_\mathrm{F} $ and $ 0 \leq \omega \leq 1 $. In contrast to \texttt{ROOK}, this metric measures the relative difference with respect to the largest energy, which compensates for the cases when the devices have uncorrelated channels but commensurable low energy.

\vspace{1mm}
\noindent \textbf{Rationale:} Intuitively, the aim of $ \mathcal{S} $ is to place in $ \mathcal{K} \backslash \mathcal{K}' $ (i.e. set of multicast-only devices) those devices that hinder more significantly the maximization of the minimum unicast SINR. This is achieved by $ f_\mathcal{S} \left( \boldsymbol{\nu} \right) $, which aims to minimize the total discordance of the co-scheduled devices. Whether such devices \emph{(i)} have highly-correlated channels among themselves or \emph{(ii)} have strongly attenuated channels and thus require high power, by not including them in $ \mathcal{K}' $, the devices in $ \mathcal{K}' $ can gain the highest profit (i.e., the minimum $ \mathrm{SINR}^\mathrm{U}_k $, $ k \in \mathcal{K}' $ is maximized). 

\subsection{Optimization of precoder and combiners} \label{precoder_combiners}
Once the scheduling variables $ \boldsymbol{\mu}^\star $ are known, we replace them in $ \mathcal{P} \left( \mathbf{W}, \mathbf{m}, \widetilde{\mathbf{B}}, \boldsymbol{\mu}^\star \right) $. Thus, the remaining problem optimizes the unicast and multicast precoders (at the gNodeB) and combiners (at the devices) as shown in
\resizebox{1.01\columnwidth}{!}{
\begin{minipage}{1.01\columnwidth}
\begin{subequations} \label{e4}
	\begin{align}
		\mathcal{D}: & \max_{
				\substack{
							\mathbf{W}, \mathbf{m}, \mathbf{B},
						 }
			   } 
		& & f_\mathcal{D} \left( \mathbf{W}, \mathbf{B} \right) \triangleq \min_{k \in \mathcal{K}'} 
								\frac{\left| \mathbf{w}^H_k \mathbf{H}_k {\mathbf{b}_k} \right|^2 }
				 				{ \sum_{ j \neq k, j \in \mathcal{K}' } \left| \mathbf{w}^H_k \mathbf{H}_k {\mathbf{b}_j} \right|^2 + \sigma^2 \left\| \mathbf{w}_k \right\|^2_2 } \nonumber		
		\\
		\vspace{-0.2cm}
		& ~~~ \mathrm{s.t.} & & \frac{\left| \mathbf{w}^H_k \mathbf{H}_k \mathbf{m} \right|^2 }
		{ \sum_{j \in \mathcal{K}'} \left| \mathbf{w}^H_k \mathbf{H}_k {\mathbf{b}_j} \right|^2 + {\sigma}^2 \left\| \mathbf{w}_k \right\|^2_2 } \geq \gamma_\mathrm{min}, \forall k \in \mathcal{K}, \nonumber
		\\
		& & & \sum_{k \in \mathcal{K}'} \left\| \mathbf{b}_k \right\|^2_2 + \left\| \mathbf{m} \right\|^2_2 \leq P_\mathrm{tx}, \nonumber
		\\ 
		& & & \left[ \mathbf{w}_k \right]_l \in \mathcal{W}, l \in \mathcal{L}, \forall k \in \mathcal{K},. \nonumber
	\end{align}
\end{subequations}
\end{minipage}
} \\

\noindent where $ \mathbf{B} = \widetilde{\mathbf{B}} \mathbf{U} $ and $ \mathbf{U} \mathbf{U}^T = \mathrm{diag} \left( \boldsymbol{\mu} \right)  $ as defined in Section \ref{system_model}. Due to coupling between $ \left\lbrace \mathbf{b}_k \right\rbrace_{k \in \mathcal{K}'} $ and $ \left\lbrace \mathbf{w}_k \right\rbrace^K_{k = 1} $, the optimization of $ \mathcal{D} $ is challenging. To cope with it, we first design the combiners $ \left\lbrace \mathbf{w}_k \right\rbrace^K_{k = 1} $ based on the channels $ \left\lbrace \mathbf{H}_k \right\rbrace^K_{k = 1} $, which are assumed to be invariant for a few channel uses. Then, we jointly optimize the unicast precoders $ \left\lbrace \mathbf{b}_k \right\rbrace_{k \in \mathcal{K}'} $ and the multicast precoder $ \mathbf{m} $.

\vspace{-2mm}
\subsubsection{Optimization of combiners $ \left\lbrace \mathbf{w}_k \right\rbrace^K_{k = 1} $} \label{subsection_design_combiners}

We define $ \mathcal{D}_1 \triangleq \cup_{k \in \mathcal{K}} \mathcal{D}_{1,k} $, where 
\begin{align} \label{e16}
	\mathcal{D}_{1,k}: \max_{ \mathbf{w}_k } \left\| \mathbf{w}^H_k \mathbf{H}_k \right\|^2_2 ~~ \mathrm{s.t.} ~~ \left| \left[ \mathbf{w}_k \right]_l \right| = \delta_\mathrm{rx}, l \in \mathcal{L}.
\end{align}

Problem $ \mathcal{D}_{1} $ designs the combiners $ \left\lbrace \mathbf{w}_k \right\rbrace^K_{k = 1} $ for all IoT devices in an independent manner. Therefore, each device can self-optimize its own combiner without need of the gNodeB. This problem admits a close-form solution that can be obtained using the Lagrange multipliers method. Specifically, the solution collapses to the principal eigenvector $ \mathbf{r}_\mathrm{max} $ of $ \mathbf{H}_k \mathbf{H}^H_k $. Then, to enforce the constant-modulus finite-resolution phase shifts, $ \mathbf{r}_\mathrm{max} $ is projected onto $ \mathcal{W} $. Therefore, for the  $ k $-th device, $ \mathbf{w}_k $ is obtained via $ \left[ \mathbf{w}_k \right]_l = \argmax_{\phi \in \mathcal{W}} \mathfrak{Re} \left\lbrace \phi^{*} \left[ \mathbf{r}_\mathrm{max} \right]_l \right\rbrace $, $ \forall l \in \mathcal{L} $. The solution of $ \mathcal{D}_{1} $ is denoted by $ \mathbf{W}^\star = \left[ \mathbf{w}_1^\star, \cdots, \mathbf{w}_K^\star \right] $.

\subsubsection{Optimization of $ \left\lbrace \mathbf{b}_k \right\rbrace_{k \in \mathcal{K}'} $ and $ \mathbf{m} $} \label{subsection_design_analog_precoder}

Assuming that $ \mathbf{g}_k = \mathbf{H}_k^H \mathbf{w}_k^\star $, the objective function of $ \mathcal{D} $ depends only on $ \mathbf{B} $. Note that $ f_\mathcal{D} \left( \mathbf{W}^\star, \mathbf{B} \right) $ is the minimum of several SINRs, which can be translated as a constraint as
\begin{subequations} \label{eB2}
	\begin{align}
		\mathcal{D}_2: & \max_{
				\substack{
							\mathbf{B}, \mathbf{m}, \alpha
						 }
			   } 
		& & \alpha \nonumber
		\\
		\vspace{-0.2cm}
		& ~~ \mathrm{s.t.} & \mathrm{R_1:} ~ & 
		\frac{ \left| \mathbf{g}^H_k {\mathbf{b}_k} \right|^2 }
		{ \sum_{ \substack{j \neq k, j \in \mathcal{K}' } } \left| \mathbf{g}^H_k {\mathbf{b}_j} \right|^2 + \sigma^2 \left\| \mathbf{w}_k^\star \right\|^2_2 } \geq \alpha, \forall k \in \mathcal{K}', \nonumber
		\\
		& & \mathrm{R_2:} ~ & \frac{\left| \mathbf{g}^H_k \mathbf{m} \right|^2 }
				{ \sum_{j \in \mathcal{K}'} \left| \mathbf{g}^H_k  {\mathbf{b}_j} \right|^2 + {\sigma}^2 \left\| \mathbf{w}_k^\star \right\|^2_2 } \geq \gamma_\mathrm{min}, \forall k \in \mathcal{K}, \nonumber
		\\
		& & \mathrm{R_3:} ~ & \textstyle \sum_{k \in \mathcal{K}'} \left\| \mathbf{b}_k \right\|^2_2 + \left\| \mathbf{m} \right\|^2_2 \leq P_\mathrm{tx}. \nonumber
	\end{align}
\end{subequations}
where $ \mathrm{R_1} - \mathrm{R_2} $ are nonconvex whereas $ \mathrm{R_3} $ is convex.

\textit{Note that $ \mathcal{D}_2 $ poses a difficulty in finding a solution as it cannot be addressed by known frameworks in its current form. In the following, we propose a reformulation of the problem that allows tailoring an algorithm to solve it. In particular, we transform $ \mathcal{D}_2 $ into a difference-of-convex (DC) programming problem, where the objective and/or constraints are convex or DC functions. Then, by harnessing the convex-concave procedure (CCP), a local optimal solution of the resulting DC programming problem can be obtained.}

\noindent \textbf{Reformulation:} With respect to $ \mathrm{R_1} $, if we bound from above the denominator with $ \sum_{ \substack{j \neq k, j \in \mathcal{K}' } } \left| \mathbf{g}^H_k {\mathbf{b}_j} \right|^2 + \sigma^2 \left\| \mathbf{w}_k^\star \right\|^2_2 \leq t_k $ and the numerator from below with $ \left| \mathbf{g}^H_k {\mathbf{b}_k} \right|^2 \geq r_k $, then $ \mathrm{R_1} $ can be equivalently rewritten as the intersection of the following constraints
\begin{equation} \nonumber
  \mathrm{R_1} =
  \begin{cases}
  	\text{ $ \mathrm{R_{1-1}}: \underbrace{r_k}_{\text{convex}} - \underbrace{\left| \mathbf{g}^H_k {\mathbf{b}_k} \right|^2}_{\text{convex}} \leq 0, \forall k \in \mathcal{K}', $} \\
   	\text{ $ \mathrm{R_{1-2}}: \underbrace{\sum_{ \substack{j \neq k, j \in \mathcal{K}' } } \left| \mathbf{g}^H_k {\mathbf{b}_j} \right|^2 + \sigma^2 \left\| \mathbf{w}_k^\star \right\|^2_2 - t_k}_{\text{convex}} \leq 0, \forall k \in \mathcal{K}', $} \\
    \text{ $ \mathrm{R_{1-3}}: \underbrace{\alpha t_k - r_k}_{\text{nonconvex}} \leq 0, \forall k \in \mathcal{K}' $}. \\ 				 	 				 
  \end{cases}
\end{equation}

In addition, we observe that the nonconvex constraint $ \mathrm{R_{1-3}} $ can be recast as
\begin{equation} \nonumber
  \mathrm{R_{1-3}}: \underbrace{\left( \alpha + t_k \right)^2 - 4 r_k}_{\text{convex}} - \underbrace{\left( \alpha - t_k \right)^2}_{\text{convex}} \leq 0, \forall k \in \mathcal{K}',
\end{equation}
which stems from the difference of squares: $ \frac{(x+y)^2 - (x-y)^2}{4} = x y $. Adopting a similar procedures as for $ \mathrm{R_1} $ reformulation, then $ \mathrm{R_2} $ can be expressed as,
\begin{equation} \nonumber
  \mathrm{R_2} =
  \begin{cases}
  	\text{ $ \mathrm{R_{2-1}}: \underbrace{p_k}_{\text{convex}} - \underbrace{\left| \mathbf{g}^H_k \mathbf{m} \right|^2}_{\text{convex}} \leq 0, \forall k \in \mathcal{K}, $} \\
   	\text{ $ \mathrm{R_{2-2}}: \underbrace{\sum_{j \in \mathcal{K}'} \left| \mathbf{g}^H_k  {\mathbf{b}_j} \right|^2 + {\sigma}^2 \left\| \mathbf{w}_k^\star \right\|^2_2 - q_k}_{\text{convex}} \leq 0, \forall k \in \mathcal{K}, $} \\
    \text{ $ \mathrm{R_{2-3}}: \underbrace{\gamma_\mathrm{min} q_k - p_k}_{\text{convex}} \leq 0, \forall k \in \mathcal{K} $}. \\ 				 	 				 
  \end{cases}
\end{equation}

Observe that $ \mathrm{R_{1-2}} $, $ \mathrm{R_{2-2}} $, $ \mathrm{R_{2-3}} $, $ \mathrm{R_{3}} $ are convex whereas $ \mathrm{R_{1-1}} $, $ \mathrm{R_{1-3}} $, $ \mathrm{R_{2-1}} $ are DC functions. Thus, with the transformations above, $ \mathcal{D}_2 $ is now a DC programming problem.

\noindent \textbf{Solution:} Optimization problems that have convex or DC objective/constraints can be efficiently tackled by means of the CCP procedure, which guarantees a stationary solution of the original problem.

\begin{mdframed} 
The CCP procedure \cite{lipp2016:variations-extensions-convex-concave-procedure} guarantees a stationary point of a DC programming problem. The main idea of CCP is to iteratively solve a sequence of convex subproblems, each of which is constructed by replacing the concave terms with first-order Taylor approximations. Consider the DC programming problem 
\begin{align}
	\mathcal{Z}: & \max_{
			\substack{
						\mathbf{z}_1, \mathbf{z}_2
				 }
  } 
	& & f \left( \mathbf{z}_1, \mathbf{z}_2 \right) \nonumber
	\\
	\vspace{-0.2cm}
	& ~~ \mathrm{s.t.} & & 
	h_i \left( \mathbf{z}_1 \right) - g_i \left( \mathbf{z}_2 \right) \leq , i = 1, \cdots, I, \nonumber
\end{align}
where $ f \left( \mathbf{z}_1, \mathbf{z}_2 \right) $ is concave in $ \mathbf{z}_1 $, $ \mathbf{z}_2 $ whereas $ h_i \left( \mathbf{z}_1 \right) $ and $ g_i \left( \mathbf{z}_2 \right) $ are convex in $ \mathbf{z}_1 $ and $ \mathbf{z}_2 $, respectively. To convexify $ \mathcal{Z} $, the concave terms, i.e. $ - g_i \left( \mathbf{z}_2 \right) $, are linearized. The resulting convexified DC programming problem is therefore expressed as
\begin{align}
	\mathcal{Z}^{(\ell)}: & \max_{
			\substack{
						\mathbf{z}_1, \mathbf{z}_2
				 }
  } 
	& & f \left( \mathbf{z}_1, \mathbf{z}_2 \right) \nonumber
	\\
	\vspace{-0.2cm}
	& ~~ \mathrm{s.t.} & & 
	h_i \left( \mathbf{z}_1 \right) - \tilde{g}_i \left( \mathbf{z}_2 \right) \leq , i = 1, \cdots, I, \nonumber
\end{align}
where $ \tilde{g}_i \left( \mathbf{z}_2 \right) = g_i \left( \mathbf{z}_2^{(\ell -1)} \right) + \nabla_{\mathbf{z}_2}^T g_i \left( \mathbf{z}_2^{(\ell -1)} \right) \left( \mathbf{z}_2 - \mathbf{z}_2^{(\ell -1)} \right) $ denotes a linearized version of $ g_i \left( \mathbf{z}_2 \right) $ around a given point $ \mathbf{z}_2^{(\ell -1)} $. Since every instance of the resulting problem $ \mathcal{Z}^{(\ell)} $ is convex, it can be solved using general-purpose solvers via interior-point methods. The process is repeated iteratively, each time refining the initial point $ \mathbf{z}_2^{(\ell -1)} \leftarrow \mathbf{z}_2 $ until a stop criterion is satisfied. Let $ N_\mathrm{conv} $ be the maximum number of iterations that $ \mathcal{Z}^{(\ell)} $ can be solved, and let $ \epsilon \geq 0 $ be a small number (e.g., $ \epsilon = 0.001 $). Thus, the iterative process stops when $ \ell = N_\mathrm{conv} $ or $ \left| f \left( \mathbf{z}_1, \mathbf{z}_2 \right) - f \left( \mathbf{z}_1^{(\ell-1)}, \mathbf{z}_2^{(\ell-1)} \right) \right| \leq \epsilon $. Further, to guarantee convergence, an initial feasible point (i.e., when $ \ell = 0 $) is required, which we discuss in Appendix \ref{appendix_A}. 
\end{mdframed} 

According to the CCP procedure described above, to solve $ \mathcal{D}_2 $, we need solve the convex problem $ \mathcal{D}_2^{(\ell)} $ iteratively until a stop criterion is met. Thus, for a given iteration $ \ell $, the convex problem $ \mathcal{D}_2^{(\ell)} $ is defined as
\resizebox{1.01\columnwidth}{!}{
\begin{minipage}{1.01\columnwidth}
\begin{subequations} \label{eB2}
	\begin{align}
		\mathcal{D}_2^{(\ell)}: & \max_{
				\substack{
							\mathbf{B}, \mathbf{m}, \alpha \\
							\mathbf{r}, \mathbf{t}, \mathbf{p}, \mathbf{q}
						 }
			   } ~ \alpha ~~ \mathrm{s.t.} ~~ \mathrm{R_{1-1}}^{(\ell)}, \mathrm{R_{1-2}}, \mathrm{R_{1-3}}^{(\ell)}, \mathrm{R_{2-1}}^{(\ell)}, \mathrm{R_{2-2}}, \mathrm{R_{2-3}}, \mathrm{R_{3}}. \nonumber
	\end{align}
\end{subequations}
\end{minipage}
} \\
\resizebox{1.01\columnwidth}{!}{
\begin{minipage}{1.01\columnwidth}
\begin{align} 
\mathrm{R_{1-1}}^{(\ell)}: r_k + \left| \mathbf{g}^H_k \mathbf{b}_k^{(l-1)} \right|^2 - 2\mathfrak{Re} \left\lbrace {\mathbf{b}_k^{(l-1)}}^H \mathbf{g}_k \mathbf{g}^H_k \mathbf{b}_k \right\rbrace \leq 0, \forall k \in \mathcal{K}', \nonumber
\end{align}
\end{minipage}
} \\
\resizebox{1.01\columnwidth}{!}{
\begin{minipage}{1.01\columnwidth}
\begin{align} 
\mathrm{R_{1-3}}^{(\ell)}: \left( \alpha + t_k \right)^2 - 4 r_k + \left( \alpha^{(l-1)} - t_k^{(l-1)} \right)^2 - ~~~~~~~~~~~~~~~~~~~ \nonumber \\ 2 \left( \alpha^{(l-1)} - t_k^{(l-1)} \right) \left( \alpha - t_k \right) \leq 0, \nonumber
\end{align}
\end{minipage}
} \\
\resizebox{1.01\columnwidth}{!}{
\begin{minipage}{1.03\columnwidth}
\begin{align} 
\mathrm{R_{2-1}}^{(\ell)}: p_k + \left| \mathbf{g}^H_k \mathbf{m}^{(l-1)} \right|^2 - 2\mathfrak{Re} \left\lbrace {\mathbf{m}^{(l-1)}}^H \mathbf{g}_k \mathbf{g}_k^H \mathbf{m} \right\rbrace \leq 0, \forall k \in \mathcal{K}. \nonumber
\end{align}
\end{minipage}
} \\

At the completion of each iteration $ \ell $, the obtained solutions $ \mathbf{B} $, $ \mathbf{m} $, $ \alpha $, $ \mathbf{t} $ are passed to $ \mathbf{B}^{(\ell)} $, $ \mathbf{m}^{(\ell)} $, $ \alpha^{(\ell)} $, $ \mathbf{t}^{(\ell)} $, which are used as the new initializations for the subsequent iteration $ \ell + 1 $. The solution of this stage is $ \mathbf{B}^\star $ and $ \mathbf{m}^\star $. For completeness, we summarize in \textit{Algorithm} the complete optimization procedure of $ \mathcal{S} $ and $ \mathcal{D} $.

\setlength{\textfloatsep}{2pt}
\captionsetup[algorithm]{labelformat=empty}
\begin{algorithm}[!t]
	\textbf{Input:} $ \left\lbrace \mathbf{H}_k \right\rbrace^K_{k = 1} $, $ \gamma_\mathrm{min} $, $ N_\mathrm{conv} $, $ \epsilon $ \\
	\textbf{Execute:} \\ 
	\vspace{0.1cm}
		\begin{tabular}{m{0.1cm} m{15cm}}
			1: & Find $ \boldsymbol{\mu}^\star $ by solving the scheduling problem $ \mathcal{S} $. \\		
			2: & Design the combiners $ \left\lbrace \mathbf{w}_k^\star  \right\rbrace^K_{k = 1} $ for all devices \\
			   & by solving problem $ {\mathcal{D}}_{1,k} $, $ \forall k \in \mathcal{K} $. \\
			3: & Design the multicast precoder $ \mathbf{m}^\star $ and the unicast \\
			   & precoders $ \left\lbrace \mathbf{b}_k^\star  \right\rbrace_{k \in \mathcal{K}'} $ by solving $ \mathcal{D}^{(\ell)}_2 $. 
		\end{tabular}
	\caption{\texttt{BEAMWAVE} optimization}
	\label{algorithm_2}
\end{algorithm}


\section{Simulation Results} \label{section_results}
\label{sec:result}

Throughout the simulations, we consider the geometric channel model defined in (\ref{e1a}), with $ L = L_1 = \cdots = L_K = 3 $ propagation paths. This assumption is compliant with the results of a measurement campaign in an industrial environment \cite{loch2019:measurement-campaign-mmwave}, where the number of propagation paths is usually between 1 to 3. The angles of arrival are uniformly distributed as $ \psi^{(l)}_k \in \left[-\pi; \pi \right] $ whereas the angles of departure are distributed as $ \phi^{(l)}_k \in \left[-\pi/3; \pi/3 \right] $. The power assigned to the combiners is $ P_\mathrm{rx} = 0 $ dBm, the noise power is $ \sigma^2 = 10 $ dBm, and the multicast QoS requirement is $ \gamma_\mathrm{min} = 4 $ ($ \sim 6 $ dB). Also, $ \omega = 0.5 $, $ N_\mathrm{conv} = 20 $ and $ \epsilon = 0.001 $. The results in this section show the average performance over $ 100 $ simulations. For the selected settings, in all the channel realizations, we have obtained feasible solutions. To solve the optimization problems, we have used \texttt{CVX}. Specifically, \texttt{CVX} and \texttt{GUROBI} were used to solve the integer linear program $ \mathcal{S} $. The convex problem $ \mathcal{D}^{(\ell)}_2 $ was solved by means of \texttt{CVX} and \texttt{SDPT3}. In the following, we examine scenarios, in which we evaluate the performance of \texttt{BEAMWAVE}.

\subsection{Minimum unicast SINR for various $ N_\mathrm{tx} $} \label{subsection_minimum_SINR_I}
\begin{figure}[!t]
\pgfplotstableread{
0	300.6234	146.5345	167.1147	167.1147	218.7725	260.3670	534.8074	310.0131	289.7304	289.7304	381.8399	497.2137
1	433.2518	225.8808	259.5697	259.5697	321.0216	386.2868	785.0557	463.0245	440.0067	440.0067	546.3746	730.8877
2	565.7604	296.8582	313.9815	313.9815	417.2531	511.1765	1024.1586	608.3798	558.9624	558.9624	723.3822	966.5721
3	213.0276	115.3772	147.3373	147.3373	172.2508	190.9780	348.9001	226.4728	224.8595	224.8595	291.0270	333.9662
4	297.9832	168.7927	215.9754	215.9754	246.1967	266.9544	515.6398	342.1055	341.0539	341.0539	434.7474	497.6390
5	356.9748	208.5988	254.3950	254.3950	288.8553	314.9828	678.2559	449.9225	437.6103	437.6103	558.4196	650.9079
6	138.7280	83.06110	102.2787	102.2787	114.7737	119.1319	234.1214	176.1848	174.9282	174.9282	216.7431	225.8926
7	197.9592	129.1385	151.8763	151.8763	171.4991	180.1531	353.7365	270.5707	271.2240	271.2240	329.9805	341.6853
8	239.4532	158.9525	185.6283	185.6283	205.2628	218.5439	464.7376	356.1628	352.5576	352.5576	429.0653	452.7688
}\scenarioIa
\pgfplotstableread{
0	352.1194	185.9239	194.2548	194.2548	255.0106	327.9352	534.8074	310.0131	289.7304	289.7304	381.8399	497.2137
1	520.3553	283.0563	313.2860	313.2860	393.0800	496.8167	785.0557	463.0245	440.0067	440.0067	546.3746	730.8877
2	684.4684	376.0084	372.4640	372.4640	513.7559	652.0559	1024.1586	608.3798	558.9624	558.9624	723.3822	966.5721
3	246.4817	139.9405	163.7721	163.7721	199.2389	226.8041	348.9001	226.4728	224.8595	224.8595	291.0270	333.9662
4	355.9615	215.1006	240.6066	240.6066	290.2559	333.7042	515.6398	342.1055	341.0539	341.0539	434.7474	497.6390
5	456.4366	276.6430	306.7875	306.7875	386.2730	437.3898	678.2559	449.9225	437.6103	437.6103	558.4196	650.9079
6	158.2934	105.6892	127.3460	127.3460	139.6710	146.6630	234.1214	176.1848	174.9282	174.9282	216.7431	225.8926
7	236.6751	168.6750	187.2718	187.2718	212.2302	223.1519	353.7365	270.5707	271.2240	271.2240	329.9805	341.6853
8	302.7940	217.4279	239.2595	239.2595	273.8294	293.1136	464.7376	356.1628	352.5576	352.5576	429.0653	452.7688
}\scenarioIb
\pgfplotstableread{
0	368.0368	195.2656	206.5614	206.5614	278.9910	345.0549	534.8074	310.0131	289.7304	289.7304	381.8399	497.2137
1	552.7169	298.1839	314.5079	314.5079	407.4810	520.4745	785.0557	463.0245	440.0067	440.0067	546.3746	730.8877
2	720.4671	395.1143	390.6305	390.6305	516.2354	674.5851	1024.1586	608.3798	558.9624	558.9624	723.3822	966.5721
3	253.7319	146.2997	168.6702	168.6702	204.9639	236.7030	348.9001	226.4728	224.8595	224.8595	291.0270	333.9662
4	372.7450	227.0329	253.6124	253.6124	316.3698	353.0172	515.6398	342.1055	341.0539	341.0539	434.7474	497.6390
5	480.5844	294.0237	319.4970	319.4970	403.6966	460.1351	678.2559	449.9225	437.6103	437.6103	558.4196	650.9079
6	165.7812	112.4664	132.5129	132.5129	145.3017	151.3810	234.1214	176.1848	174.9282	174.9282	216.7431	225.8926
7	253.7044	179.1533	200.1509	200.1509	233.0732	241.9651	353.7365	270.5707	271.2240	271.2240	329.9805	341.6853
8	323.6531	235.2380	252.1174	252.1174	297.3342	315.1765	464.7376	356.1628	352.5576	352.5576	429.0653	452.7688
}\scenarioIc
\pgfplotstableread{
0	369.1300	197.0286	208.7138	208.7138	280.4756	345.8501	534.8074	310.0131	289.7304	289.7304	381.8399	497.2137
1	558.9612	301.1982	327.5272	327.5272	410.3707	531.7828	785.0557	463.0245	440.0067	440.0067	546.3746	730.8877
2	733.7278	399.5926	399.7434	399.7434	529.3219	692.5373	1024.1586	608.3798	558.9624	558.9624	723.3822	966.5721
3	252.9622	147.3331	169.8164	169.8164	204.8116	233.9525	348.9001	226.4728	224.8595	224.8595	291.0270	333.9662
4	376.5507	228.5136	257.1341	257.1341	311.4127	354.6576	515.6398	342.1055	341.0539	341.0539	434.7474	497.6390
5	484.3001	297.4524	325.4943	325.4943	397.1765	463.9873	678.2559	449.9225	437.6103	437.6103	558.4196	650.9079
6	167.2023	112.4810	133.5501	133.5501	144.1134	152.8065	234.1214	176.1848	174.9282	174.9282	216.7431	225.8926
7	253.1808	181.8690	201.1726	201.1726	229.9317	240.0552	353.7365	270.5707	271.2240	271.2240	329.9805	341.6853
8	325.9089	236.3563	254.0027	254.0027	292.9779	312.3865	464.7376	356.1628	352.5576	352.5576	429.0653	452.7688
}\scenarioId

	\begin{tikzpicture}[thick,scale=0.63, every node/.style={scale=0.95}, inner sep=0.25mm]
	\draw [fill=gray, opacity=0.15, draw=gray!0] (0.0,0) rectangle +(4.3,3.42);
	\draw [fill=gray, opacity=0.15, draw=gray!0] (8.6,0) rectangle +(4.3,3.42);
	\draw (2, 3.2) node {\scriptsize $K' = 3$};	
	\draw (6.5, 3.2) node {\scriptsize $K' = 4$};	
	\draw (11, 3.2) node {\scriptsize $K' = 5$};	
	\begin{axis}[ybar,
        width = 14.5cm,
        height = 5cm,
        ymin = 0,
        ymax = 1100, 
        bar width = 4pt, 
        enlarge x limits = 0.06,      
        ylabel = {Minimum unicast SINR},
        xlabel = {Number of antennas at the gNodeB},
        y label style={align=center, font=\small},
        x label style={align=center, font=\small},
        xtick = data,
        xticklabels = {\strut $N_\mathrm{tx} = 16$, \strut $N_\mathrm{tx} = 24$, \strut $N_\mathrm{tx} = 36$, \strut $N_\mathrm{tx} = 16$, \strut $N_\mathrm{tx} = 24$, \strut $N_\mathrm{tx} = 36$,\strut $N_\mathrm{tx} = 16$, \strut $N_\mathrm{tx} = 24$, \strut $N_\mathrm{tx} = 36$},
        xticklabel style={yshift=1ex},		
        legend columns = 6,
        legend style={at={(-0.00,1.08)},anchor=south west, font=\footnotesize, fill = none, /tikz/every even column/.append style={column sep=0.55mm}},
        legend cell align={left}
        ]
        
		\addplot[draw=black, fill=dcolor1, every node near coord/.append style={xshift=-1pt, yshift=0.6pt}] table[x index=0,y index=7] \scenarioIa;	
		\addlegendentry{\texttt{XHAUS}}
		\addplot[draw=black, fill=dcolor2, every node near coord/.append style={xshift=-1pt, yshift=0.3pt}] table[x index=0,y index=8] \scenarioIa; 	
		\addlegendentry{\texttt{RANDOM}};	
		\addplot[draw=black, fill=dcolor3, every node near coord/.append style={xshift=-1pt, yshift=0pt}] table[x index=0,y index=9] \scenarioIa;	
		\addlegendentry{\texttt{BEAMWAVE-CORR}};
		\addplot[draw=black, fill=dcolor4, every node near coord/.append style={xshift=-1pt, yshift=0pt}] table[x index=0,y index=10] \scenarioIa;	
		\addlegendentry{\texttt{BEAMWAVE-PAWN}}; 
		\addplot[draw=black, fill=dcolor5, every node near coord/.append style={xshift=-1pt, yshift=-0.5pt}] table[x index=0,y index=11] \scenarioIa;	
		\addlegendentry{\texttt{BEAMWAVE-ROOK}}; 
		\addplot[draw=black, fill=dcolor6, every node near coord/.append style={xshift=-1pt, yshift=-0.9pt}] table[x index=0,y index=12] \scenarioIa;	
		\addlegendentry{\texttt{BEAMWAVE-KING}}; 		

	\end{axis}

	\end{tikzpicture}
	\centering
	\vspace{-1mm}
	\caption{Achievable minimum unicast SINR for varying $ N_{\mathrm{tx}} $ at the gNodeB.}
	\label{figure_unicast_minimum_SINR_I}
\end{figure}
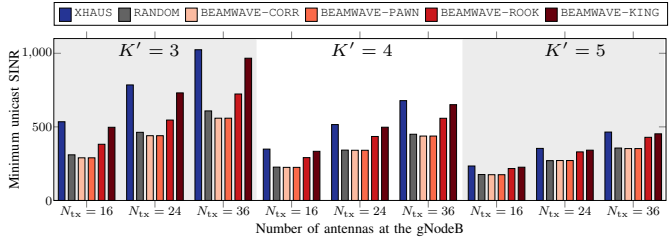

Fig. \ref{figure_unicast_minimum_SINR_I} depicts the impact of different $ N_\mathrm{tx} $ configurations on the minimum unicast SINR when the total number of devices in the system is $ K = 6 $ and the number of \emph{dual-layer} devices $ K' = \left\lbrace 3, 4, 5 \right\rbrace $ varies. In this case, we have assumed that the IoT devices are equipped with a single antenna, i.e., $ N_\mathrm{rx} = 1 $ and the gNodeB can transmit with a maximum power $ P_\mathrm{tx} = 35 $ dBm. 

As a general trend, we observe that increasing the number of \emph{dual-layer} devices $ K'$ decreases the minimum unicast SINR. This occurs because the limited power $ P_\mathrm{tx} $ at the gNodeB is divided into a greater number of scheduled devices, thus reducing the individual allocation of power for each \textit{dual-layer} device. Also, serving more \emph{dual-layer} devices translates to producing more interference, thus impacting the SINR. On the contrary, increasing $ N_\mathrm{tx} $ improves the minimum unicast SINR. Essentially, a larger $ N_\mathrm{tx} $ reduces the beamwidth that can be produced by the antenna array at the gNodeB, thus allowing to form more directional transmissions with reduced interference. 
\begin{figure*}[!t]
\pgfplotstableread{
1	167.2023	112.481		133.5501	133.5501	144.1134	152.8065	234.1214	176.1848	174.9282	174.9282	216.7431	225.8926
2	253.2162	163.2053	182.6108	192.7991	219.7526	231.8748	341.7406	257.4891	254.2640	256.1209	312.9889	330.6381
3	297.590		186.751		208.0887	220.7256	240.9281	258.6671	412.0458	304.3863	302.9182	309.5313	367.8746	390.5638
4	344.8526	209.9108	228.2728	262.412		281.3068	303.6404	480.5359	346.7345	343.2814	361.7527	428.7061	453.6331
}\scenarioIIIa
\pgfplotstableread{
1	167.2023	112.4810 	133.5501	133.5501	144.1134	152.8065	234.1214	176.1848	174.9282	174.9282	216.7431	225.8926
2	269.7888	175.1344	199.3252	210.3786	227.3955	245.5574	375.6383	283.0272	280.1065	283.1858	346.0618	367.7373
3	351.1249	218.3198	244.8537	261.4258	288.6952	307.3697	488.7542	357.6878	357.5822	366.2691	438.7012	469.9602
4	436.8283	278.2850 	294.4102	346.9411	363.5010	375.2508	604.8468	450.5734	442.9663	471.7755	551.6605	583.3865
}\scenarioIIIb
\pgfplotstableread{
1	167.2023	112.4810	133.5501	133.5501	144.1134	152.8065	234.1214	176.1848	174.9282	174.9282	216.7431	225.8926
2	279.5630	178.8735	204.7429	212.6554	233.9247	251.4673	385.4824	288.8826	286.4407	290.0066	352.7815	374.9458
3	373.2998	231.4311	265.3334	279.9457	311.0444	328.5296	519.0768	379.4211	378.7131	389.0630	466.4074	499.6366
4	471.3419	291.3435	315.9266	370.6609	388.4506	408.2091	645.3132	479.1450	471.4916	505.2042	589.5817	624.5821
}\scenarioIIIc
\pgfplotstableread{
1	167.2023	112.4810	133.5501	133.5501	144.1134	152.8065	234.1214	176.1848	174.9282	174.9282	216.7431	225.8926
2	276.0692	179.4030	198.6930	214.5385	229.3994	244.0851	389.5462	292.1860	288.9247	292.3915	357.2677	379.1618
3	377.5436	232.0985	267.2140	281.1105	309.8677	327.1432	527.3587	384.8702	382.8408	393.3982	473.7911	507.3613
4	477.5958	301.6310	319.5432	381.1028	398.5149	421.3143	657.4101	487.8843	479.7472	514.3687	601.0253	637.0501
}\scenarioIIId

\begin{subfigure}[t]{0.239\textwidth}
	\begin{tikzpicture}[thick, scale=0.75, every node/.style={scale=0.96}, inner sep=0.25mm]
		\begin{axis}[
			width = 6.5cm,
        	height = 5cm,
			ylabel = {Minimum unicast SINR},
        	y label style={align=center, font=\small},
			ymajorgrids,
			xlabel = {Number of antennas at each IoT device},
        	x label style={align=center, font=\small},
        	xtick = data,
        	xticklabels = {\strut $N_\mathrm{rx} = 1$, \strut $N_\mathrm{rx} = 2$, \strut $N_\mathrm{rx} = 3$, \strut $N_\mathrm{rx} = 4$},
			enlarge x limits = 0.15,
			ymin=100,ymax=675,
			legend columns = 6,
			legend style={at={(0.7,1.04)}, anchor=south west, font=\footnotesize, fill = none, /tikz/every even column/.append style={column sep=0.05cm}},
		]
			
			\addplot[color = dcolor1, mark = pentagon*, mark options = {scale = 0.9, fill = dcolor1, solid}, line width = 0.3pt, smooth] table [x=0, y=7]{\scenarioIIIa};
			\addlegendentry{\texttt{XHAUS}};
			
			\addplot[color = dcolor2, mark = square*, mark options = {scale = 0.7, fill = dcolor2, solid}, line width = 0.3pt, smooth] table [x=0, y=8]{\scenarioIIIa};
			\addlegendentry{\texttt{RANDOM}};
			
			\addplot[color = dcolor3, mark = triangle*, mark options = {scale = 0.9, fill = dcolor3, solid}, line width = 0.3pt, smooth] table [x=0, y=9]{\scenarioIIIa};
			\addlegendentry{\texttt{BEAMWAVE-CORR}};
			
			\addplot[color = dcolor4, mark = diamond*, mark options = {scale = 1, fill = dcolor4, solid}, line width = 0.3pt, smooth] table [x=0, y=10]{\scenarioIIIa};
			\addlegendentry{\texttt{BEAMWAVE-PAWN}};
			
			\addplot[color = dcolor5, mark = *, mark options = {scale = 0.8, fill = dcolor5, solid}, line width = 0.3pt, smooth] table [x=0, y=11]{\scenarioIIIa};
			\addlegendentry{\texttt{BEAMWAVE-ROOK}};
			
			\addplot[color = dcolor6, mark = triangle*, mark options = {scale = 0.9, fill = dcolor6, solid,rotate=180}, line width = 0.3pt, smooth] table [x=0, y=12]{\scenarioIIIa};
			\addlegendentry{\texttt{BEAMWAVE-KING}};
		\end{axis}
	\end{tikzpicture}
	\caption{$ L_\mathrm{rx} = 2 $}
	\label{figure_unicast_minimum_SINR_IIA}
\end{subfigure}
\hspace{1mm}
\begin{subfigure}[t]{0.239\textwidth}
	\begin{tikzpicture}[thick,scale=0.75, every node/.style={scale=0.96}, inner sep=0.25mm]
		\begin{axis}[
			width = 6.5cm,
        	height = 5cm,
			ylabel = {Minimum unicast SINR},
        	y label style={align=center, font=\small},
			ymajorgrids,
			xlabel = {Number of antennas at IoT device},
        	x label style={align=center, font=\small},
        	xtick = data,
        	xticklabels = {\strut $N_\mathrm{rx} = 1$, \strut $N_\mathrm{rx} = 2$, \strut $N_\mathrm{rx} = 3$, \strut $N_\mathrm{rx} = 4$},
			enlarge x limits = 0.15,
			ymin=100,ymax=675,
			legend columns = 5,
			legend style={at={(-0.12,1.08)}, anchor=south west, font=\normalsize, fill = none, /tikz/every even column/.append style={column sep=0.046cm}},
		]

			\addplot[color = dcolor1, mark = pentagon*, mark options = {scale = 0.9, fill = dcolor1, solid}, line width = 0.3pt, smooth] table [x=0, y=7]{\scenarioIIIb};
			
			\addplot[color = dcolor2, mark = square*, mark options = {scale = 0.7, fill = dcolor2, solid}, line width = 0.3pt, smooth] table [x=0, y=8]{\scenarioIIIb};
			
			\addplot[color = dcolor3, mark = triangle*, mark options = {scale = 0.9, fill = dcolor3, solid}, line width = 0.3pt, smooth] table [x=0, y=9]{\scenarioIIIb};
			
			\addplot[color = dcolor4, mark = diamond*, mark options = {scale = 1, fill = dcolor4, solid}, line width = 0.3pt, smooth] table [x=0, y=10]{\scenarioIIIb};
			
			\addplot[color = dcolor5, mark = *, mark options = {scale = 0.8, fill = dcolor5, solid}, line width = 0.3pt, smooth] table [x=0, y=11]{\scenarioIIIb};
			
			\addplot[color = dcolor6, mark = triangle*, mark options = {scale = 0.9, fill = dcolor6, solid,rotate=180}, line width = 0.3pt, smooth] table [x=0, y=12]{\scenarioIIIb};
		\end{axis}
	\end{tikzpicture}
	\caption{$ L_\mathrm{rx} = 4 $}
	\label{figure_unicast_minimum_SINR_IIB}
\end{subfigure}
\hspace{1mm}
\begin{subfigure}[t]{0.239\textwidth}
	\begin{tikzpicture}[thick,scale=0.75, every node/.style={scale=0.96}, inner sep=0.25mm]
		\begin{axis}[
			width = 6.5cm,
        	height = 5cm,
			ylabel = {Minimum unicast SINR},
        	y label style={align=center, font=\small},
			ymajorgrids,
			xlabel = {Number of antennas at IoT device},
        	x label style={align=center, font=\small},
        	xtick = data,
        	xticklabels = {\strut $N_\mathrm{rx} = 1$, \strut $N_\mathrm{rx} = 2$, \strut $N_\mathrm{rx} = 3$, \strut $N_\mathrm{rx} = 4$},
			enlarge x limits = 0.15,
			ymin=100,ymax=675,
			legend columns = 6,
			legend style={at={(-0.12,1.08)}, anchor=south west, font=\normalsize, fill = none, /tikz/every even column/.append style={column sep=0.046cm}},
		]
		
			\addplot[color = dcolor1, mark = pentagon*, mark options = {scale = 0.9, fill = dcolor1, solid}, line width = 0.3pt, smooth] table [x=0, y=7]{\scenarioIIIc};
			
			\addplot[color = dcolor2, mark = square*, mark options = {scale = 0.7, fill = dcolor2, solid}, line width = 0.3pt, smooth] table [x=0, y=8]{\scenarioIIIc};
			
			\addplot[color = dcolor3, mark = triangle*, mark options = {scale = 0.9, fill = dcolor3, solid}, line width = 0.3pt, smooth] table [x=0, y=9]{\scenarioIIIc};
			
			\addplot[color = dcolor4, mark = diamond*, mark options = {scale = 1, fill = dcolor4, solid}, line width = 0.3pt, smooth] table [x=0, y=10]{\scenarioIIIc};
			
			\addplot[color = dcolor5, mark = *, mark options = {scale = 0.8, fill = dcolor5, solid}, line width = 0.3pt, smooth] table [x=0, y=11]{\scenarioIIIc};
			
			\addplot[color = dcolor6, mark = triangle*, mark options = {scale = 0.9, fill = dcolor6, solid,rotate=180}, line width = 0.3pt, smooth] table [x=0, y=12]{\scenarioIIIc};
		\end{axis}
	\end{tikzpicture}
	\caption{$ L_\mathrm{rx} = 8 $}
	\label{figure_unicast_minimum_SINR_IIC}
\end{subfigure}
\hspace{1mm}
\begin{subfigure}[t]{0.239\textwidth}
	\begin{tikzpicture}[thick,scale=0.75, every node/.style={scale=0.96}, inner sep=0.25mm]
		\begin{axis}[
			width = 6.5cm,
        	height = 5cm,
			ylabel = {Minimum unicast SINR},
        	y label style={align=center, font=\small},
			ymajorgrids,
			xlabel = {Number of antennas at IoT device},
        	x label style={align=center, font=\small},
        	xtick = data,
        	xticklabels = {\strut $N_\mathrm{rx} = 1$, \strut $N_\mathrm{rx} = 2$, \strut $N_\mathrm{rx} = 3$, \strut $N_\mathrm{rx} = 4$},
			enlarge x limits = 0.15,
			ymin=100,ymax=675,
			legend columns = 5,
			legend style={at={(-0.12,1.08)}, anchor=south west, font=\normalsize, fill = none, /tikz/every even column/.append style={column sep=0.046cm}},
		]
			
			\addplot[color = dcolor1, mark = pentagon*, mark options = {scale = 0.9, fill = dcolor1, solid}, line width = 0.3pt, smooth] table [x=0, y=7]{\scenarioIIId};
			
			\addplot[color = dcolor2, mark = square*, mark options = {scale = 0.7, fill = dcolor2, solid}, line width = 0.3pt, smooth] table [x=0, y=8]{\scenarioIIId};
			
			\addplot[color = dcolor3, mark = triangle*, mark options = {scale = 0.9, fill = dcolor3, solid}, line width = 0.3pt, smooth] table [x=0, y=9]{\scenarioIIId};
			
			\addplot[color = dcolor4, mark = diamond*, mark options = {scale = 1, fill = dcolor4, solid}, line width = 0.3pt, smooth] table [x=0, y=10]{\scenarioIIId};
			
			\addplot[color = dcolor5, mark = *, mark options = {scale = 0.8, fill = dcolor5, solid}, line width = 0.3pt, smooth] table [x=0, y=11]{\scenarioIIId};
			
			\addplot[color = dcolor6, mark = triangle*, mark options = {scale = 0.9, fill = dcolor6, solid,rotate=180}, line width = 0.3pt, smooth] table [x=0, y=12]{\scenarioIIId};
		\end{axis}
	\end{tikzpicture}
	\caption{$ L_\mathrm{rx} = 16 $}
	\label{figure_unicast_minimum_SINR_IID}
\end{subfigure}
\caption{Achievable minimum SINR for varying $N_{\mathrm{rx}}$ and $ L_\mathrm{rx} $ at each IoT device.}
\label{figure_unicast_minimum_SINR_II}
\end{figure*}
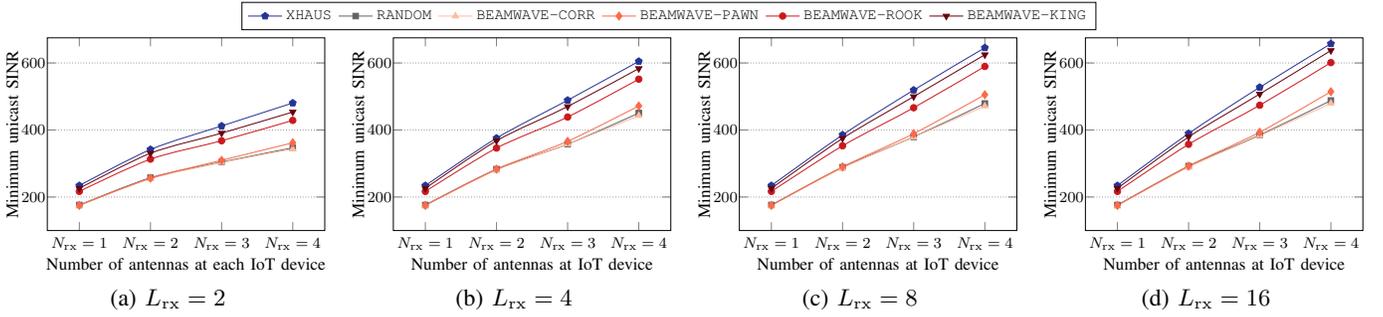

Another general trend in Fig. \ref{figure_unicast_minimum_SINR_I} is that \texttt{XHAUS} (exhaustive search) exhibits the highest performance in all configurations as it schedules the optimum subset of $ K' $ \emph{dual-layer} devices. By leveraging the channel correlation, \texttt{BEAMWAVE-CORR}\footnote{As mentioned in Section \ref{scheduling}, when $ N_\mathrm{rx} = 1 $, \texttt{PAWN} and \texttt{CORR} result in the same value. For this reason, we observe that \texttt{BEAMWAVE-CORR} and \texttt{BEAMWAVE-PAWN} attain the same performance.} only performs slightly better than \texttt{RANDOM}. Thus, scheduling decisions based solely on the channel correlation are insufficient to devise an optimal scheduler for LDM systems. On the contrary, \texttt{BEAMWAVE-ROOK} and \texttt{BEAMWAVE-KING}, which additionally include channel energy information, clearly outperform \texttt{RANDOM}. These two schemes achieve up to $ 60.38\% $ and $ 77.68\% $ higher SINR, respectively, compared to \texttt{RANDOM}. Noteworthily, throughout all the results in Fig. \ref{figure_unicast_minimum_SINR_I}, \texttt{BEAMWAVE-ROOK} and \texttt{BEAMWAVE-KING} perform at worst $ 30.4\% $ and $ 14.13\%$ below \texttt{XHAUS}, respectively.

\subsection{Minimum unicast SINR for various $ N_\mathrm{rx} $} \label{subsection_minimum_SINR_II}

Fig. \ref{figure_unicast_minimum_SINR_II} shows the impact of varying $ N_\mathrm{rx} $ and $ L_\mathrm{rx} $ on the minimum unicast SINR when $ K = 6 $, $ K' = 5 $, $ N_\mathrm{tx} = 16 $, and $ P_\mathrm{tx} = 35 $ dBm. In this setting, the IoT devices have a single RF chain that is connected to $ N_\mathrm{rx} $ antennas. As a result, the devices are not capable of implementing any type of linear processing for interference mitigation but can perform constrained beamsteering due to constraint $ \mathrm{C_4} $ in $ \mathcal{P} $.

In all subfigures in Fig. \ref{figure_unicast_minimum_SINR_II}, we observe that the minimum unicast SINR improves as the number of receive antennas increases. With larger $ N_\mathrm{rx} $, the devices can shape more directional reception patterns to mitigate undesired signals. In particular, up to $ 60\% $ gain can be achieved with $ L_\mathrm{rx} = 4 $ when varying $ N_\mathrm{tx} $ from $ 1 $ to $ 2 $. Also, since augmenting $ L_\mathrm{rx} $ results in higher-resolution phase shifts, we observe performance improvement through Fig. \ref{figure_unicast_minimum_SINR_IIA} to Fig. \ref{figure_unicast_minimum_SINR_IID}. In particular, gains up to $ 16.00\% $, $ 30.70\% $ and $ 49.47\% $ are achieved when increasing $ L_\mathrm{rx} $ from $ 2 $ to $ 4 $, $ 4 $ to $ 8 $ and $ 8 $ to $ 16 $, respectively. 

By comparing the performance of the proposed scheduling schemes under all assessed settings, the scheme that attains superior performance is \texttt{BEAMWAVE-KING}. In particular, \texttt{BEAMWAVE-KING} is outperformed by at most $ 5.60\% $ when compared to the optimal highly complex \texttt{XHAUS}. 

\subsection{Spectral efficiency} \label{s7c}
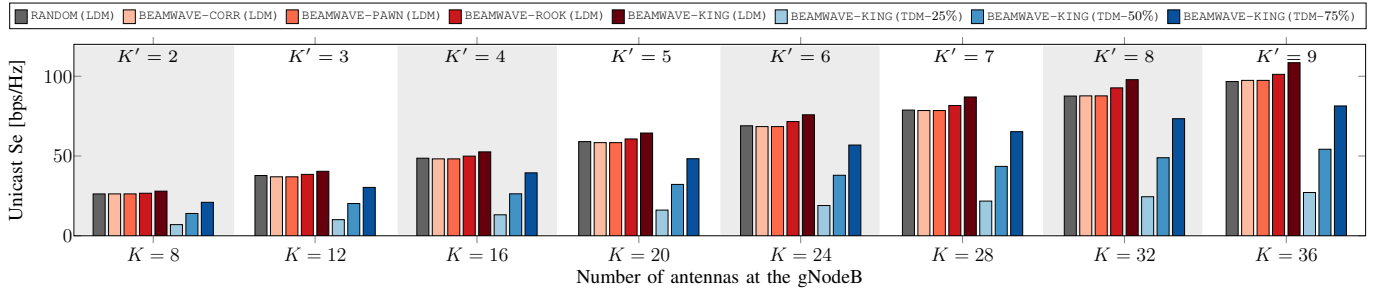
\begin{figure*}[!t]
\pgfplotstableread{
0	24.4286   24.5569   24.5569   25.2434   26.2621    6.5655   13.1310   19.6965   26.2317   26.2606   26.2606   26.6625   27.9722    6.9930   13.9861   20.9791
1	34.6045   34.4505   34.4505   36.0448   37.8029    9.4507   18.9014   28.3522   37.7742   36.9736   36.9736   38.4896   40.4279   10.1070   20.2140   30.3209
2   44.1900   44.8682   44.8682   46.7346   49.0810   12.2702   24.5405   36.8107   48.6301   48.2116   48.2116   49.9276   52.5983   13.1496   26.2992   39.4487
3   52.5242   54.3474   54.3474   56.3195   59.5254   14.8813   29.7627   44.6440   59.0446   58.3847   58.3847   60.6914   64.4103   16.1026   32.2051   48.3077
4   60.6363   63.7002   63.7002   66.3746   69.7482   17.4370   34.8741   52.3111   68.9533   68.4360   68.4360   71.6085   75.8406   18.9601   37.9203   56.8804
5   67.8570   72.9518   72.9518   74.6533   78.8162   19.7040   39.4081   59.1121   78.8051   78.5139   78.5139   81.6559   87.0071   21.7518   43.5035   65.2553
6   72.2935   81.1461   81.1461   84.2410   87.0645   21.7661   43.5322   65.2983   87.5876   87.7022   87.7022   92.7204   97.8583   24.4646   48.9292   73.3938
7   77.6837   89.9609   89.9609   91.1252   96.5248   24.1312   48.2624   72.3936   96.6862   97.4173   97.4173  101.1993  108.5098   27.1275   54.2549   81.3824
}\scenarioIVa

	\begin{tikzpicture}[thick,scale=0.65, every node/.style={scale=0.95}, inner sep=0.25mm]
	\draw [fill=gray, opacity=0.15, draw=gray!0] (0.0,0) rectangle +(3.3,3.91);
	\draw [fill=gray, opacity=0.15, draw=gray!0] (6.6,0) rectangle +(3.3,3.91);
	\draw [fill=gray, opacity=0.15, draw=gray!0] (13.2,0) rectangle +(3.3,3.91);
	\draw [fill=gray, opacity=0.15, draw=gray!0] (19.8,0) rectangle +(3.3,3.91);
	\draw (1.5, 3.7) node {\scriptsize $K' = 2$};
	\draw (5, 3.7) node {\scriptsize $K' = 3$};	
	\draw (8.2, 3.7) node {\scriptsize $K' = 4$};	
	\draw (11.6, 3.7) node {\scriptsize $K' = 5$};
	\draw (14.85, 3.7) node {\scriptsize $K' = 6$};
	\draw (18.2, 3.7) node {\scriptsize $K' = 7$};
	\draw (21.47, 3.7) node {\scriptsize $K' = 8$};
	\draw (24.8, 3.7) node {\scriptsize $K' = 9$};
	\begin{axis}[ybar,
        width = 28cm,
        height = 5.5cm,
        ymin = 0,
        ymax = 120, 
        bar width = 7pt, 
        enlarge x limits = 0.07,      
        ylabel = {Unicast Se [bps/Hz]},
        xlabel = {Number of antennas at the gNodeB},
        y label style={align=center, yshift=2ex, font=\large},
        x label style={align=center, font=\large},
        xtick = data,
        xticklabels = {\strut $K = 8$, \strut $K = 12$, \strut $K = 16$, \strut $K = 20$, \strut $K = 24$,\strut $K = 28$, \strut $K = 32$, \strut $K = 36$, \strut $K = 40$, \strut $K = 44$, \strut $K = 48$, \strut $K = 52$},
        xticklabel style={yshift=0ex, font=\large},		
        yticklabel style={xshift=0ex, font=\large},
        legend columns = 8,
        legend style={at={(-0.05,1.08)}, anchor=south west, align=left, fill = none, /tikz/every even column/.append style={column sep=0.55mm}, font=\footnotesize},
        legend cell align={left}
        ]

		\addplot[draw=black, fill=dcolor2, every node near coord/.append style={xshift=-1pt, yshift=0.3pt}] table[x index=0,y index=9] \scenarioIVa; 	
		\addlegendentry{\texttt{RANDOM(LDM)}};	
		\addplot[draw=black, fill=dcolor3, every node near coord/.append style={xshift=-1pt, yshift=0pt}] table[x index=0,y index=10] \scenarioIVa;	
		\addlegendentry{\texttt{BEAMWAVE-CORR(LDM)}};
		\addplot[draw=black, fill=dcolor4, every node near coord/.append style={xshift=-1pt, yshift=0pt}] table[x index=0,y index=11] \scenarioIVa;	
		\addlegendentry{\texttt{BEAMWAVE-PAWN(LDM)}}; 
		\addplot[draw=black, fill=dcolor5, every node near coord/.append style={xshift=-1pt, yshift=-0.5pt}] table[x index=0,y index=12] \scenarioIVa;	
		\addlegendentry{\texttt{BEAMWAVE-ROOK(LDM)}}; 
		\addplot[draw=black, fill=dcolor6, every node near coord/.append style={xshift=-1pt, yshift=-0.9pt}] table[x index=0,y index=13] \scenarioIVa;	
		\addlegendentry{\texttt{BEAMWAVE-KING(LDM)}}; 
		\addplot[draw=black, fill=dcolor7, every node near coord/.append style={xshift=-1pt, yshift=-0.9pt}] table[x index=0,y index=14] \scenarioIVa;	
		\addlegendentry{\texttt{BEAMWAVE-KING(TDM-$25\%$)}}; 
		\addplot[draw=black, fill=dcolor8, every node near coord/.append style={xshift=-1pt, yshift=-0.9pt}] table[x index=0,y index=15] \scenarioIVa;	
		\addlegendentry{\texttt{BEAMWAVE-KING(TDM-$50\%$)}}; 
		\addplot[draw=black, fill=dcolor9, every node near coord/.append style={xshift=-1pt, yshift=-0.9pt}] table[x index=0,y index=16] \scenarioIVa;	
		\addlegendentry{\texttt{BEAMWAVE-KING(TDM-$75\%$)}}; 		

	\end{axis}
	
	\end{tikzpicture}
	\centering
\caption{Spectral efficiency performance for varying $ K $ and $ K' = K / 4 $.}
\label{figure_unicast_spectral_efficiency_I}
\end{figure*}

In this scenario we consider $ N_\mathrm{rx} = 1 $, $ N_\mathrm{tx} = 32 $, , $ P_\mathrm{tx} = 45 $ dBm and a varying number of devices $ K = \left\lbrace 8, 12, 16, 20, 24, 28, 32, 36 \right\rbrace $. In particular, the number of scheduled \emph{dual-layer} devices changes according to $ K’ = K / 4 $. In Fig. \ref{figure_unicast_spectral_efficiency_I}, we show the unicast spectral efficiency (SE) attained by \texttt{BEAMWAVE}. Due to the exponential growth in the number of scheduling combinations, the results with \texttt{XHAUS} are not presented in this scenario. However, \texttt{BEAMWAVE-KING} is taken as reference as it was shown in previous scenarios that its performance is at most $ 14.13\% $ below the optimality of \texttt{XHAUS}. Further, we also use our proposed scheduler with T/FDM systems. 

Note that while \texttt{RANDOM} scheduling performs as equally well as \texttt{BEAMWAVE} for small $ K’ $ (since the generated interference is low), we observe that when $ K’ $ is large (e.g., $ K' = 16 $) there is a significant performance gap. This shows that scheduling exerts a critical task, specially in LDM systems which generate additional inter-layer interference between unicast and multicast signals. Besides, we observe that LDM outperforms TDM, where the time allotted for unicast transmissions is $ 25 \% $, $ 50 \% $ and $ 75 \% $ of the total available\footnote{In the TDM case, the IoT devices are served in two time windows. In the first window, with duration $ T_m $, all IoT devices in $ \mathcal{K} $ are served with the multicast control signal. In the second window, with duration $ T_u $, a subset of devices $ \mathcal{K}' $ are served with unicast signals (e.g., software updates), such that $ T_m + T_u = 1 $. In our simulations, we have varied $ T_u = \left\lbrace 0.25, 0.50, 0.75 \right\rbrace $.  }. The remaining time is used for transmitting the multicast signal. Specifically, in the TDM case, we have also used \texttt{BEAMWAVE} to make the selection of unicast devices that yields the \emph{max-min} SINR.

\subsection{Computational complexity} \label{subsection_complexity}

In Table \ref{table_computational_complexity}, we show the complexity of the benchmarked schemes. In particular, $ \mathcal{C}_\mathcal{S} $ is the complexity of the proposed scheduler, where $ M = \frac{K(K+1)}{2} $ is the number of $ 0-1 $ variables and $ C = K^2 -K + 1 $ is the number of constraints. As a reference, we have used the runtime of Vaidya's algorithm for the linear program, which \texttt{GUROBI} solves via the branch and bound (BnB) procedure. The complexity $ \mathcal{C}_{{\mathcal{D}}_{1,k}} $ stems from the singular value decomposition (SVD) used to obtain the principal eigenvector, as described in Section \ref{subsection_design_combiners}. Also, $ \mathcal{C}_{\mathcal{D}_2} $ is derived based on the complexity required by interior point methods. Finally, $ \mathcal{C}_\texttt{XHAUS} $, $ \mathcal{C}_\texttt{BEAMWAVE} $ and $ \mathcal{C}_\texttt{RANDOM} $ denote the overall complexities of the schemes \texttt{XHAUS}, \texttt{BEAMWAVE} and \texttt{RANDOM} respectively.
\begin{table}[h!]
	\footnotesize
	\caption{Computational complexity}
	\centering
	\begin{tabular}{|l|l|}
		\hline
		\centering
		{\bf \centering{Notation}} & {\bf \centering Complexity }	\\ 
		\hline
		\hline
		$ \mathcal{C}_\mathcal{S} $ & $ \mathcal{O}\left( 2^M (M+C)^{1.5} M \right) $ \\ 
		\hline		
		$ \mathcal{C}_{{\mathcal{D}}_{1,k}} $ & $ \mathcal{O}\left( N_\mathrm{rx}^{3} \right) $ \\	
		\hline
		$ \mathcal{C}_{\mathcal{D}_2} $ & $ N_\mathrm{conv} \cdot \mathcal{O} \left( \left( N_\mathrm{tx} K' \left( K + K' \right) \right)^{3.5} \right) $ \\	
		\hline
		$ \mathcal{C}_\texttt{RANDOM} $ & $ \mathcal{C}_{\mathcal{D}_2} + K \cdot \mathcal{C}_{{\mathcal{D}}_{1,k}} $ \\
		\hline
		$ \mathcal{C}_\texttt{BEAMWAVE} $ & $ \mathcal{C}_{\mathcal{D}_2} + K \cdot \mathcal{C}_{{\mathcal{D}}_{1,k}} + \mathcal{C}_\mathcal{S} $ \\
		\hline
		$ \mathcal{C}_\texttt{XHAUS} $ & $ {K \choose K'} \cdot \mathcal{C}_{\mathcal{D}_2} + K \cdot \mathcal{C}_{{\mathcal{D}}_{1,k}} $ \\
		\hline
	\end{tabular}
	\label{table_computational_complexity}
\end{table}

\section{Conclusions} \label{s9}

In this paper we investigated the cross-layer optimization of beamforming and scheduling for mmWave LDM systems, aiming to support future Industry 4.0 scenarios. In particular, through the adoption of LDM, multiple signal layers can be transmitted simultaneously using the same radio resources. For smart factory settings, we assumed that a superior-importance safety/control multicast message is required to be ubiquitous to all the devices in the system. In addition, due to insufficient RF chains, inferior-importance private unicast information is simultaneously transmitted to a selected group of scheduled devices with the aim of maximizing the minimum SINR. Due to NP-hardness of the problem, we proposed \texttt{BEAMWAVE} which partitions the problem into \emph{(i)} beamforming and \emph{(ii)} scheduling. For device scheduling, we proposed a novel formulation, where we devised three metrics based on channel features, namely \texttt{PAWN}, \texttt{ROOK}, and \texttt{KING} to guide the selection decision. Further, we designed a precoder (i.e., transmit beamformer) with remarkable performance adopting the convex-concave procedure. We showed that our proposed scheme attains high spectral efficiency and outperforms orthogonal multiplexing schemes such as T/FDM.

\section*{Acknowledgment} \label{section_acknowledgment}
The research is in part funded by the Deutsche Forschungsgemeinschaft (DFG) within the B5G-Cell project in SFB 1053 MAKI and by the LOEWE initiative (Hesse, Germany) within the emergenCITY center.

\bibliographystyle{IEEEtran}
\bibliography{ref}

\begin{appendices}

\setcounter{equation}{0}
\renewcommand{\theequation}{C.\arabic{equation}}
\section{Initial feasible point for $ \mathcal{D}_2 $ } \label{appendix_A}

In order to find $ \mathbf{B}^{(0)} $, $ \mathbf{m}^{(0)} $, $ \alpha^{(0)} $, $ \mathbf{t}^{(0)} $ we proceed as follows. First, let us define $ a $ as the power of the multicast precoder $ \mathbf{m} $, such that $ \mathbf{m} = \sqrt{a} \widehat{\mathbf{m}} $, $ \left\| \widehat{\mathbf{m}} \right\|_2^2 = 1 $. Similarly, we define  $ a_k $ as the power of the unicast precoder $ \mathbf{b}_k $, $ k \in \mathcal{K}' $ such that  $ \mathbf{b}_k = \sqrt{a}_k \widehat{\mathbf{b}}_k $, $ \left\| \widehat{\mathbf{b}}_k \right\|_2^2 = 1 $. Now, we let $ \left\lbrace \widehat{\mathbf{b}}_k \right\rbrace_{k \in \mathcal{K}'} $ be the zero-forcing precoders \cite{hao2017:energy-efficiency-mmwave-noma}. On the other hand, we let $ \widehat{\mathbf{m}} $ be the principal eigenvector of the aggregate channels of all users. Thus, we define
\begin{subequations} \label{eB2}
	\begin{align}
		\mathcal{D}_2^\mathrm{ini}: & \min_{
				\substack{
							\left\lbrace a_k \right\rbrace_{k \in \mathcal{K}'}, a
						 }
			   } 
		& & \sum_{k \in \mathcal{K}'} \sum_{ \substack{j \neq k, j \in \mathcal{K}' } } a_j \left| h_{k,j} \right|^2  \nonumber
		\\
		\vspace{-0.2cm}
		& ~~~~~~ \mathrm{s.t.} & & \frac{a \left| h_k \right|^2 }
				{ \sum_{j \in \mathcal{K}'} a_j \left| h_{k,j} \right|^2 + {\sigma}^2 \left\| \mathbf{w}_k^\star \right\|^2_2 } \geq \gamma_\mathrm{min}, \forall k \in \mathcal{K}, \nonumber
		\\
		& & & \textstyle \sum_{k \in \mathcal{K}'} a_k \left\| \widehat{\mathbf{b}}_k \right\|^2_2 +  a \left\| \widehat{\mathbf{m}} \right\|^2_2 \leq P_\mathrm{tx}, \nonumber
	\end{align}
\end{subequations}
where $ h_{k,j} = \mathbf{g}^H_k \widehat{\mathbf{b}}_j $ and $ h_k = \mathbf{g}^H_k \mathbf{m} $. Note that $ \mathcal{D}_2^\mathrm{ini} $ is a linear programming problem. Also, observe that any feasible solution for $ \mathcal{D}_2^\mathrm{ini} $ will be feasible for $ \mathcal{D}_2 $. In particular, the objective function of $ \mathcal{D}_2^\mathrm{ini} $ minimizes the total unicast interference perceived by all IoT devices (i.e., sum of all terms in the denominator of $ \mathrm{R_1} $ in $ \mathcal{D}_2 $). Once $ \mathcal{D}_2^\mathrm{ini} $ is solved, we obtain a solution $ \left( \left\lbrace a_k^\star \right\rbrace_{k \in \mathcal{K}'}, a^\star \right) $. Harnessing this outcome, we obtain the initial feasible points for $ \mathcal{D}_2^{(0)} $ by defining $ \mathbf{b}_k^{(0)} = a_k^\star \widehat{\mathbf{b}}_k $, $ \mathbf{m}^{(0)} = a^\star \widehat{\mathbf{m}} $, $ t_k^{(0)} = \sum_{ \substack{j \neq k, j \in \mathcal{K}' } } a_j^\star \left| h_{k,j} \right|^2 + \sigma^2 \left\| \mathbf{w}_k^\star \right\|^2_2 $, $ \alpha^{(0)} = \min_{k \in \mathcal{K}'} \frac{ a_k^\star \left| h_{k,k} \right|^2 } { \sum_{ \substack{j \neq k, j \in \mathcal{K}' } } a_j^\star \left| h_{k,j} \right|^2 + \sigma^2 \left\| \mathbf{w}_k^\star \right\|^2_2 } $. 
\end{appendices}

\end{document}